\documentclass{emulateapj}
\usepackage{txfonts}
\usepackage[figuresright]{rotating}

\def\hub{\ifmmode H_\circ\else H$_\circ$\fi}

\shorttitle{Properties of M33 globular clusters} \shortauthors{Ma}

\begin{document}
\slugcomment{AJ, in press}
\title{STRUCTURAL PARAMETERS FOR 10 HALO GLOBULAR CLUSTERS IN M33}

\author{Jun Ma,\altaffilmark{1}}

\altaffiltext{1}{Key Laboratory of Optical Astronomy, National Astronomical Observatories, Chinese Academy of Sciences, Beijing 100012, China; majun@nao.cas.cn}


\begin{abstract}
In this paper, we present the properties of 10 halo globular clusters with luminosities $L\simeq 5-7\times 10^5{L_\odot}$ in the Local Group galaxy M33 using the images of {\it Hubble Space Telescope} Wide Field Planetary Camera 2 in the F555W and F814W bands. We obtained ellipticities, position angles and surface brightness profiles for them. In general, the ellipticities of M33 sample clusters are similar to those of M31 clusters. The structural and dynamical parameters are derived by fitting the profiles to three different models combined with mass-to-light ratios ($M/L$ values) from population-synthesis models. The structural parameters include core radii, concentration, half-light radii and central surface brightness. The dynamical parameters include the integrated cluster mass, integrated binding energy, central surface mass density and predicted line-of-sight velocity dispersion at the cluster center. The velocity dispersions of four clusters predicted here agree well with the observed dispersions by Larsen et al. The results here showed that the majority of the sample halo globular clusters are well fitted by King model as well as by Wilson model, and better than by S\'{e}rsic model. In general, the properties of clusters in M33, M31 and the Milky Way fall in the same regions of parameter spaces. The tight correlations of cluster properties indicate a ``fundamental plane'' for clusters, which reflects some universal physical conditions and processes operating at the epoch of cluster formation.
\end{abstract}

\keywords{galaxies: individual (M33) -- galaxies: star clusters: general -- galaxies: stellar content}

\section{Introduction}
\label{s:intro}
Globular clusters (GCs) are effective laboratories for studying stellar evolution and stellar dynamics. They are ancient building blocks of galaxies which can help us understand the formation and evolution of their parent galaxies. In addition, GCs exhibit surprisingly uniform properties, suggesting a common formation mechanism. It is well known that studying the spatial structures and dynamics of GCs is of great importance for understanding both their formation condition and dynamical evolution within the tidal fields of their galaxies \citep{barmby07}.

Structural and dynamical parameters of GCs as described by surface brightness and velocity profiles can be fitted by a number of different models. The models include those of the empirical, single-mass, modified isothermal spheres \citep{king62,king66,wilson75}, the isotropic multi-mass \citep{df76}, the anisotropic multi-mass \citep{gg79,meylan88,meylan89}, and the power-law surface-density profiles \citep{sersic68,elson87}. Since the pioneer work of \citet{mm05}, three models are often used in the fits. First is based on single-mass, isotropic, modified isothermal sphere developed by \citet{king66}. Second is a further modification of a single-mass, isotropic isothermal sphere based on the ad hoc stellar distribution function of \citet{wilson75}. The third model is based on the $R^{1/n}$ surface density profile of \citet{sersic68}. Using the three models, some authors have achieved some success in determining structural and dynamical parameters of clusters in the Local galaxies: the Milky Way, the Large and Small Magellanic Clouds, Fornax and Sagittarius dwarf spheroidal galaxies \citep{mm05}, M31 \citep{barmby07,barmby09,wang13}, and NGC 5128 \citep{mclau08}.

Due to their proximity, galaxies in the Local Group provide us with ideal targets for detailed studies of spatial structures and dynamics of GCs. The GCs of the Milky Way and M31 have received close attention \citep[see][and references there]{mm05,wang13}. However, before \citet{roman12} determined structural parameters for 161 star clusters in M33, structural studies of the star cluster system of M33 are very limited. \citet{CBF99a,CBF99b} derived core radii of 60 star clusters in M33 using linear correlations with the measured full width at half-maximum (FWHM) of each cluster using the {\it Hubble Space Telescope} ({\it HST}) images, and found that core radii of M33 star clusters are smaller on average than those of Galactic GCs and LMC populous clusters at similar magnitudes. \citet{larsen02} determined structural parameters for four halo GCs in M33 by fitting King model \citep{king66} to the surface brightness of the {\it HST} images. These authors presented that the four M33 clusters have similar structural parameters, and fall in the same ``fundamental plane'' and binding energy-luminosity relations as Milky Way and M31 clusters. \citet{Robert11} measured structural parameters of six outer clusters of M33 by fitting the models of \citet{king62} and \citet{king66}, and of \citet{wilson75}, using the Canada-France-Hawaii Telescope (CFHT)/MegaCam data as part of the Pan-Andromeda Archaeological Survey. The structural parameters obtained by \citet{Robert11} include the concentration parameter, and core, half-light and tidal radii. \citet{roman12} presented detailed morphological properties of M33 star clusters using the {\it HST} images. They derived ellipcities and position angles for 161 star clusters, and found that M33 clusters are more flattened than those of the Milky Way and M31. They concluded that the cluster flattening of M33 is resulted from tidal forces. \citet{roman12} also derived structural parameters for these star clusters including core radii, concentration, half-light radii and central surface brightness by fitting the empirical King model \citep{king62} and Elson-Freeman-Fall model \citep{elson87}. The results of \citet{roman12} showed differences in the structural evolution between the M33 cluster system and clusters in nearby galaxies. In this paper, we will determine structural and dynamical parameters for 10 halo GCs in M33 by fitting three structural models \citep{king66,wilson75,sersic68} to their surface brightness profiles, and compare the resulting fundamental planes of GC parameters in different galaxies. Through out this work a distance to M33 of 847 kpc [$(m-M)_0 = 24.64$] \citep{Galleti04} has been adopted. At that distance, 1 arcsec corresponds 4.11 pc.

This paper is organized as follows. In Section 2, we present the data-processing steps to derive the surface brightness profiles. In Section 3, we determine structural and dynamical parameters of the clusters and make some comparisons with previous studies. A discussion on the correlations of the derived parameters is given in Section 4. Finally, we summarize our results in Section 5.

\section{DATA AND ANALYSIS PROCEDURES}
\label{data.sec}

\subsection{Halo Globular Cluster Sample in M33}

The sample of M33 halo GCs in this paper is from \citet{sara98}, who originally selected 10 star clusters in M33  based on their halo-like kinematics and red colors ($B-V >0.6$). \citet{sara98} considered that these clusters should be as close an analogy as possible to the halo clusters in the Milky Way. The observations used in the present work come from the {\it HST} programs 5914, of which the sample star clusters were observed by the {\it HST}/Wide Field Planetary Camera 2 (WFPC2) in F555W and F814W bands. Each cluster was centered in the PC1 chip, and was observed with a total exposure time of 4800 s in F555W band and 5200 s in F814W band \citep[see][for details]{sara00}. We obtained the combined drizzled images from the Hubble Legacy Archive. Figure 1 shows images of the sample halo GCs in M33 observed in the F555W and F814W filters of WFPC2/{\it HST}. We want to point out that \citet{larsen02} have determined structural parameters for four of the sample clusters by fitting King model \citep{king66} to the surface brightness of the same {\it HST} images as those used in this paper. However, we still selected them as our sample of GCs because we want to study them in the way as \citet{mm05} did for the GCs in the Milky Way, \citet{barmby07}, \citet{barmby09} and \citet{wang13} did for the GCs in M31, and \citet{mclau08} did for the GCs in NGC 5128. We will determine not only structural parameters, but also dynamical parameters for the sample clusters. In addition, we will study correlations between the parameters obtained here combined with those obtained by \citet{mm05} for Milky Way GCs, and by \citet{barmby07} and \citet{wang13} for M31 GCs (see Section 4). Especially, \citet{larsen02} measured the velocity dispersions for four sample clusters using high-dispersion spectra from the HIRES echelle spectrograph on the Keck I telescope. So, we can compare the velocity dispersions predicted here with those measured by \citet{larsen02} to check our results. In addition, another sample halo GC (U137) had been studied by \citet{CBF99a,CBF99b} and \citet{roman12} using the {\it HST} images different from those used here. \citet{CBF99a,CBF99b} derived core radii of U137 using linear correlations with the FWHM. \citet{roman12} derived structural parameters of U137 including core radii, concentration, half-light radii and central surface brightness.

\begin{figure*}
\figurenum{1}
\includegraphics[width=20.0cm]{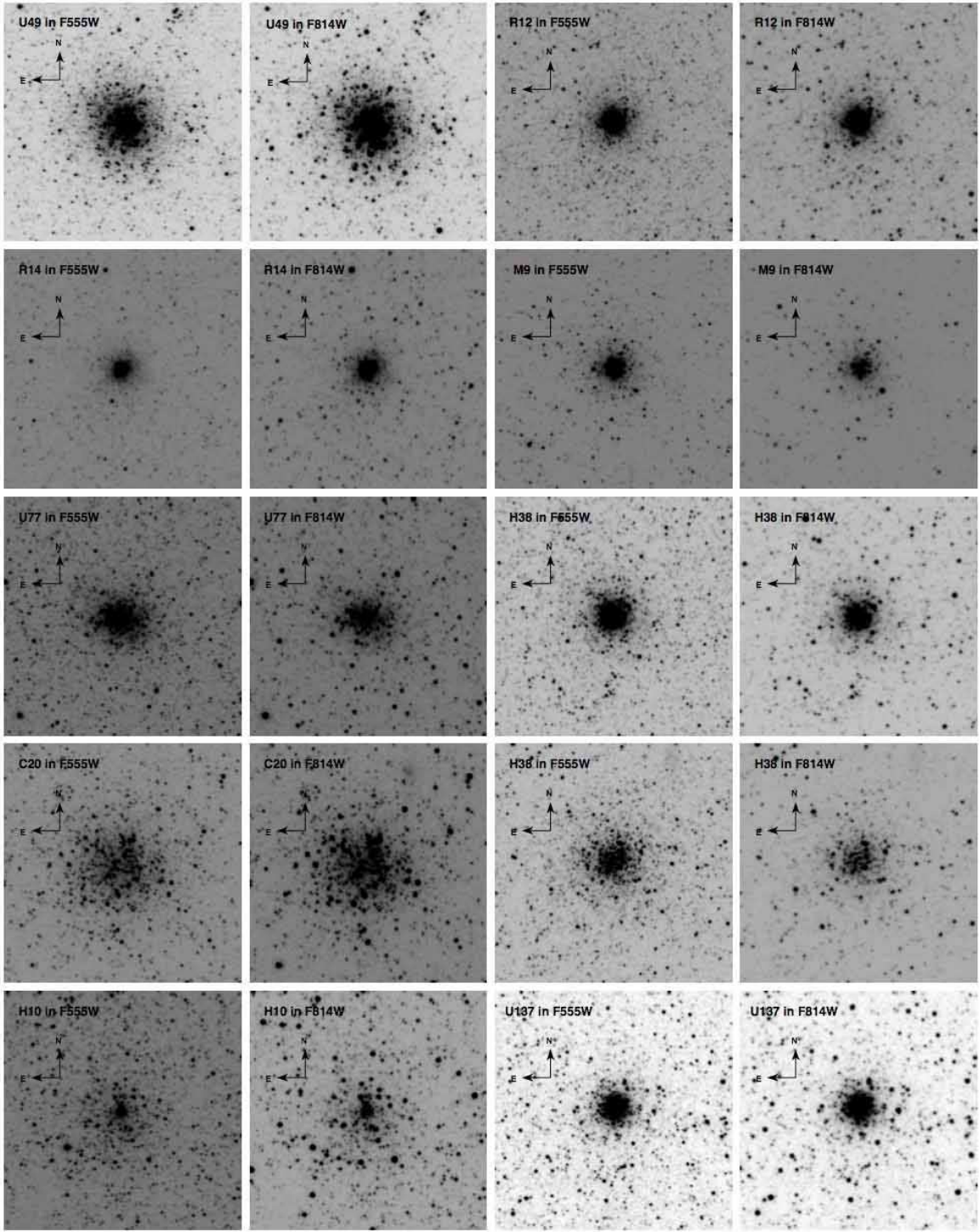}
\vspace{-4.0cm} \caption{Images of halo GCs in M33 observed in the F555W and F814W filters of WFPC2/{\it HST}. The image
size is $11.25'' \times 11.25''$ for each panel.} \label{fig1}
\end{figure*}

When we calculate the mass-to-light ratios ($M/L$ values) which are used to derive the dynamical parameters, the ages and metallicities are used. In this paper, we used the ages and metallicities obtained by \citet{Ma02d} and \citet{Ma04a}. \citet{Ma02d} estimated the ages of these M33 halo GCs based on the integrated spectral energy distributions (SEDs) obtained by the Beijing--Arizona--Taipei--Connecticut (BATC) system with 13 intermediate-band filters. These authors fitted the SEDs with the simple stellar population (SSP) model to determine the ages of star clusters. \citet{CBF02} also estimated the ages of these clusters by comparing the integrated photometry of them with the SSP model in the color--color diagram. \citet{Ma04a} showed that the ages estimated by \citet{Ma02d} are in good agreement with those estimated by \citet{CBF02} (see Table 1 of Ma et al. 2004 for details). \citet{PP09} estimated ages and metallicities of 100 star clusters in M33 including four sample star clusters by fitting the theoretical isochrones to the observational color-magnitude diagrams. In general, the ages obtained by \citet{PP09} are younger than those in \citet{CBF02} and \citet{Ma02d}. For the metallicities of the sample star clusters, some works \citep{Cohen84,CS88,bh91,sara98,sara00,Ma04a} determined them. Using two reddening-independent techniques, \citet{Cohen84} determined the metallicities for four sample clusters. Using the integrated spectra, \citet{CS88} estimated the metallicities for seven sample clusters, and \citet{bh91} estimated the metallicities for five sample clusters, respectively. \citet{sara98,sara00} estimated the metallicities for these 10 sample star clusters based on the shape and color of the red giant branch obtained using the images of the {\sl HST}/WFPC2. \citet{Ma04a} estimated the metallicities of 31 M33 old star clusters including these 10 sample clusters based on the SSP model and the photometries of the BATC intermediate-band filters. From Table 5 of \citet{Ma04a} and their discussions, we can see that the results of \citet{Ma04a} are consistent with those published by previous studies. The reddening values of nine sample clusters were obtained by \citet{sara98} except for R14. For R14, \citet{Ma02d} determined its reddening value. In this paper, we adopted the reddening values of the sample clusters obtained by \citet{sara98} and \citet{Ma02d}. The $BVI$ magnitudes of eight sample clusters are from \citet{ma12} and \citet{ma13}, who presented new $UBVRI$ photometry for 626 star clusters and candidates in M33 including eight sample clusters (except for R12 and R14) using archival images from the Local Group Galaxies Survey \citep{massey06}. The $BVI$ magnitudes of R12 and R14 are from \citet{CS88}. Galactocentric distances of the sample clusters from \citet{sara98} are adopted here. The parameters of the sample star clusters are listed in Table 1, which will be used in following sections.

\subsection{Ellipticity, Position Angle, and Surface Brightness Profile}
\label{brightness.sec}

The data analysis procedures to measure surface brightness profiles of clusters have been described in \citet{barmby07}. Here we briefly summarize the procedures. Surface photometries of each cluster were obtained from the drizzled images, using the {\sc iraf} task {\sc ellipse}. The center position of a sample cluster was fixed at a value derived by object locator of {\sc ellipse} task, however an initial center position was determined by centroiding. Elliptical isophotes were fitted to the data, with no sigma clipping. We ran two passes of {\sc ellipse} task, the first pass was run in the usual way, with ellipticity and position angle allowed to vary with the isophote semimajor axis. In the second pass, surface brightness profiles on fixed, zero-ellipticity isophotes were measured, since we choose to fit circular models for the intrinsic cluster structure and the point spread function (PSF) as \citet{barmby07} did. In order to derive required smooth profiles by ELLIPSE, we filtered the images of the sample clusters with an median filter except for R12 and R14, since the profiles of these two clusters are smooth enough. As \citet{roman12} adopted, we filtered the images of the sample clusters with an $11\times11$ $\rm{pixel^2}$ median filter except for H10 and C20. For C20, we filtered its images with an $20\times20$ $\rm{pixel^2}$ median filter, and for H10, we filtered its images with an $30\times30$ $\rm{pixel^2}$ median filter, in order to derive smooth profiles of these two clusters. Figures 2 and 3 plot the ellipticity ($\epsilon=1-b/a$) and position angle (P.A.) as a function of the semimajor axis ($a$) in the F555W and F814W bands for the sample clusters. The errors were generated by the {\sc iraf} task {\sc ellipse}, in which the ellipticity errors were obtained from the internal errors in the harmonic fit, after removal of the first and second fitted harmonics. Figure 2 shows that the ellipticities are generally large at small radii for most sample clusters. Figure 3 shows that the position angles are occasionally wildly varying for some sample clusters. This is likely to be produced by internal errors in the {\sc ellipse}. In the far outer parts of the clusters, the ellipticities and the position angles are poorly constrained due to the low signal-to-noise ratio. It is true that the ellipticities are very different between small radii and large radii for most sample clusters. So, the final ellipticity and position angle for each cluster were calculated as the average of the values between 0.2 and 1.5 arcsec (indicate by dashed lines in Figures 2 and 3) obtained in the first pass of {\sc ellipse}, where the quantities are more stable. Table 1 lists the average ellipticity and position angle for every star cluster. Errors correspond to the standard deviation of the mean. In addition, the ellipticities in the different bands are very different at small radii, since the random fluctuations due to individual stars make the fits meaningless at small radii.

Raw output from package {\sc ellipse} is in terms of counts s$^{-1}$ pixel$^{-1}$, which needs to multiply by $494=(1~ {\rm pixel}/0.045~{\rm arcsec})^2$ to convert to counts s$^{-1}$ ${\rm arcsec}^{-2}$. A formula is used to transform counts to surface brightness in magnitude calibrated on the {\sc vegamag} system,

\begin{equation}
\mu/{\rm mag~arcsec^{-2}=
-2.5 \log(counts~s^{-1}~arcsec^{-2}) + Zeropoint}.
\end{equation}

As noted by \citet{barmby07}, occasional oversubtraction of background during the multidrizzling in the automatic reduction pipeline leads to ``negative'' counts in some pixels, so we worked in terms of linear intensity instead of surface brightness in magnitude. With updated absolute magnitudes of the sun $M_{\odot}$ from Table 2 of \citet{wang13}, the equation for transforming counts to surface brightness in intensity is derived,

\begin{equation}
I/L_{\odot}~{\rm pc^{-2}
\simeq Conversion~Factor\times(counts~s^{-1}~arcsec^{-2})}.
\end{equation}

Converting from luminosity density in $L_{\odot}~{\rm pc^{-2}}$ to surface brightness in magnitude was done according to
\begin{equation}
\mu/{\rm mag~arcsec^{-2}}=
-2.5\log(I/L_{\odot}~{\rm pc^{-2}}) + {\rm Coefficient}.
\end{equation}

The Zeropoints, Conversion Factors, and Coefficients used in these transformations for each filter are from Table 2 of \citet{wang13}. In this paper, the final, calibrated intensity profiles for the sample GCs but with no extinction corrected are listed in Table 2. The reported intensities are calibrated on the {\sc vegamag} scale. In Table 2, column 6 gives a flag for each point, which has the same meaning as \citet{barmby07} and \citet{mclau08} defined. The points flagged with``OK'' are used to constrain the model fit, while the points flagged with ``DEP'' are those that may lead to excessive weighting of the central regions of clusters \citep[see][for details]{barmby07,mclau08}. In addition, points marked with ``BAD'' are those individual isophotes that deviated strongly from their neighbors or showed irregular features, which were deleted by hand.

\begin{figure*}[!htb]
\figurenum{2}
\center
\resizebox{\width}{!}{\rotatebox{-90}
{\includegraphics[width=0.75\textwidth]{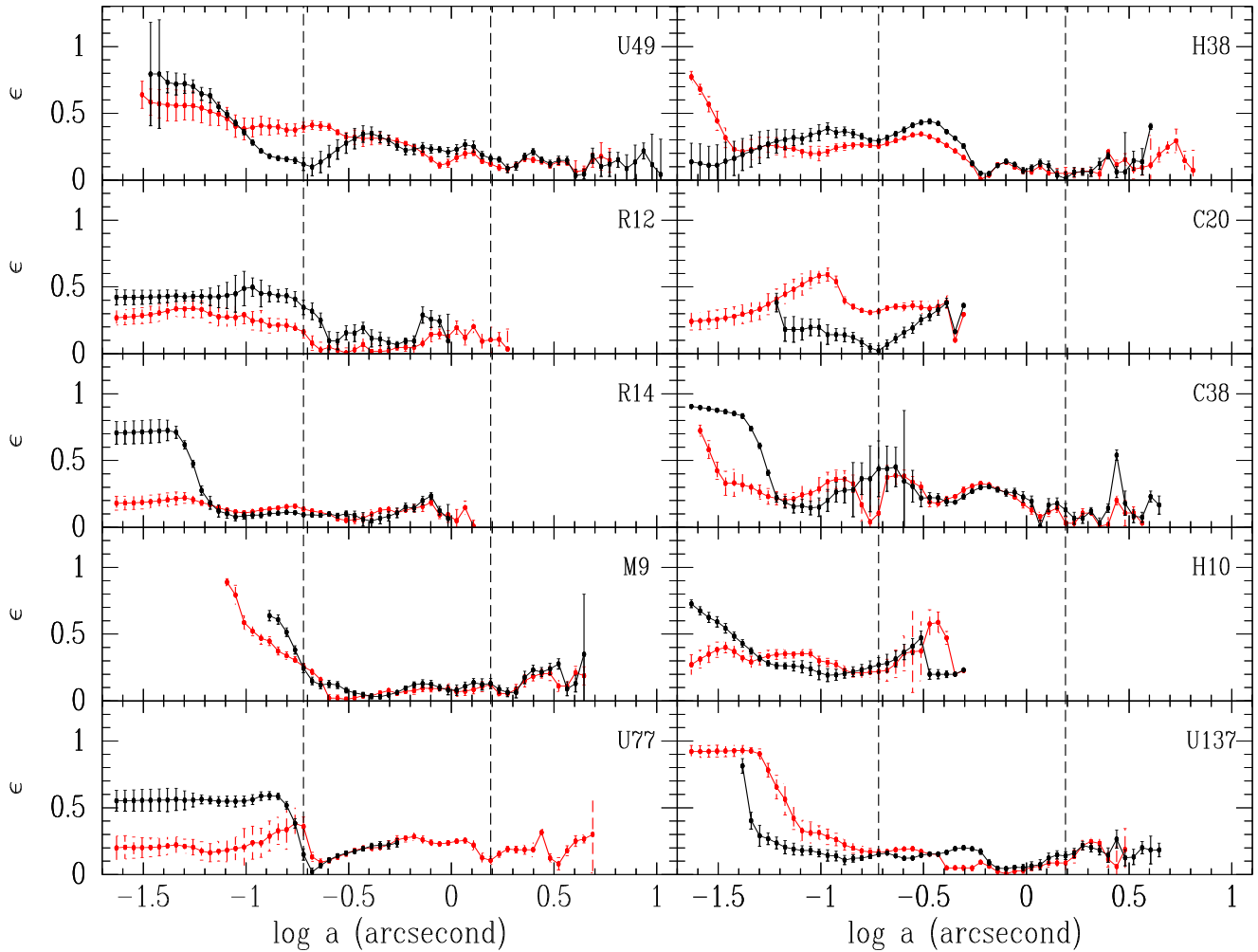}}}
\caption{Ellipticity ($\epsilon$) as a function of the semimajor axis ($a$) in the F555W (red dots) and F814W (black dots) filters of WFPC2/{\it HST} for each sample cluster. Dashed lines indicate the points of 0.2 and 1.5 arcsec along the semimajor axis.}
\label{fig2}
\end{figure*}

\begin{figure*}[!htb]
\figurenum{3}
\center
\resizebox{\width}{!}{\rotatebox{-90}
{\includegraphics[width=0.75\textwidth]{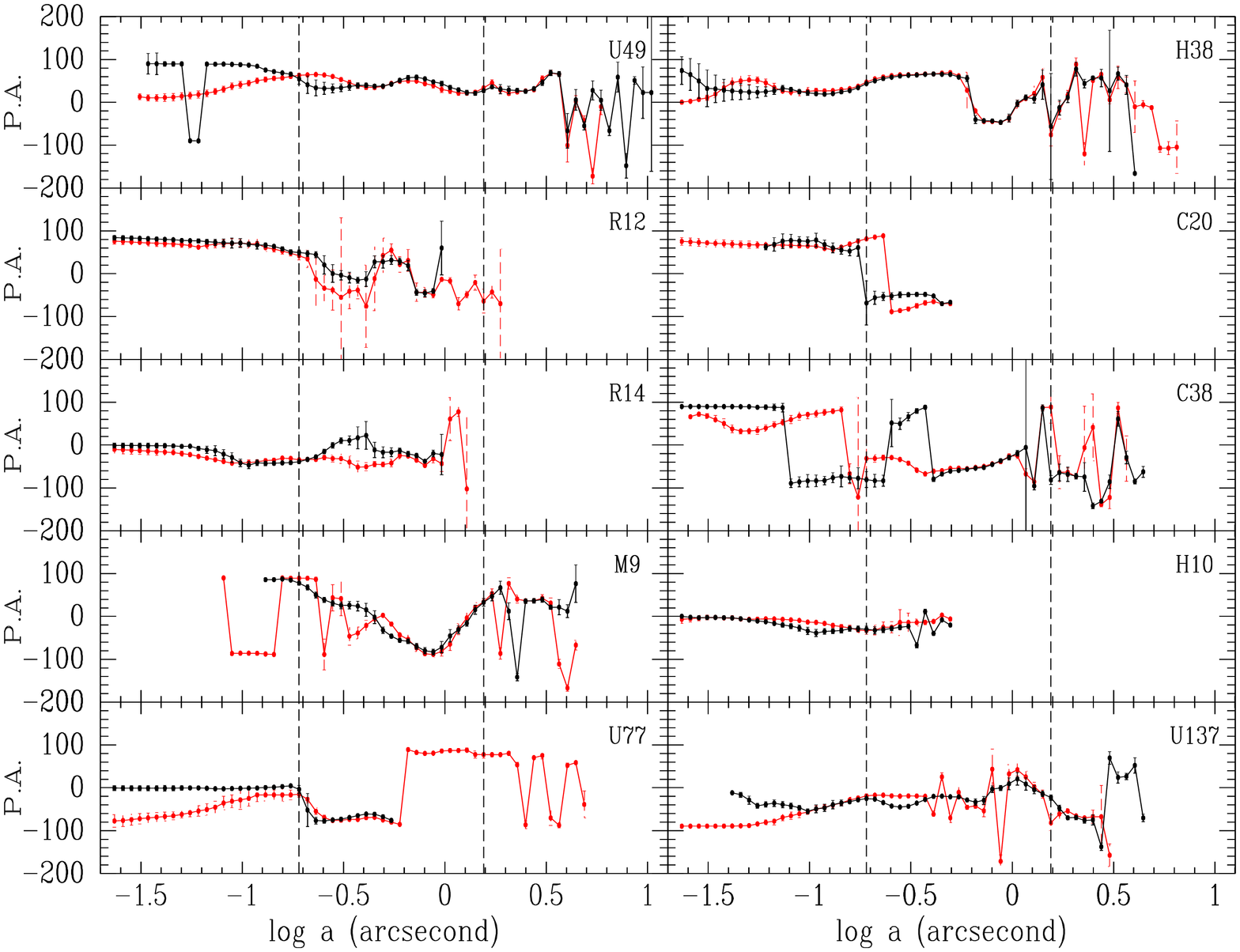}}}
\caption{Position angle (P.A.) as a function of the semimajor axis ($a$) in the F555W (red dots) and F814W (black dots)
filters of WFPC2/{\sl HST} for each sample cluster. Dashed lines indicate the points of 0.2 and 1.5 arcsec along the semimajor axis.}
\label{fig3}
\end{figure*}

\subsection{Point-spread Function}

The PSF models are critical to accurately measure the shapes of objects in images taken with {\sl HST} \citep{rhodes06}. In this paper, we chose not to deconvolve the data, instead fitting structural models after convolving them with a simple analytic description of the PSF as \citet{barmby07} and \citet{wang13} did for M31 star clusters \citep[see][for details]{barmby07,wang13}. A simple analytic description of the PSFs for the WFPC2 bands has been given by \citet{wang13}, which will be adopted here.

\subsection{Extinction and Magnitude Transformation}

When we fit models to the surface brightness profiles of the sample clusters, we will correct the intensity profiles for extinction. The effective wavelengths of the WFPC2 F555W and F814W bands are $\lambda_{\rm eff}\simeq$ 5439.0 and 8012.2 {\AA} \citep{siri05}. With the extinction curve $A_{\lambda}$ taken from \citet{car89} with $R_V=3.1$, two formulas for computing ${A_{\rm F555W}}$ and ${A_{\rm F814W}}$ are derived: ${A_{\rm F555W}}\simeq2.81~E_{B-V}$, and ${A_{\rm F814W}}\simeq1.85~E_{B-V}$ \citep[also see Table 2 of][]{wang13}. In addition, for easy comparison with catalogs of the GCs in the Milky Way (see Section 4 for details), we transform the WFPC2 magnitudes in the F555W band to the standard $V$. \citet{siri05} has given transformations from WFPC2 to standard $BVRI$ magnitudes both on observed and synthetic methods (see their Table 22). As synthetic transformations are based on larger color range and more safely employed, they should be considered the norm, unless some indicated cases \citep{siri05}. We used the synthetic transformation from F555W to $V$ magnitude on the {\sc vegamag} scale with a quadratic dependence on dereddened $(V-I)_0$. With the magnitudes in $V$ and $I$ bands and reddening values listed in Table 1, we found the $(V-I)_0$ values of all the sample clusters are larger than 0.4. So, the following transformation formula from \citet{holtzman95} was applied here,
\begin{equation}
(V-{\rm F555W})_0=0.006-0.05(V-I)_0+0.009(V-I)_0^2,
\end{equation}
for which we estimated a precision of about $\pm 0.05$ mag.

\section{MODEL FITTING}
\label{fit.sec}

\subsection{Structure Models}

As \citet{barmby07}, \citet{mclau08}, and \citet{wang13} did, we used three structural models to fit star cluster surface profiles. These models are developed by \citet{king66}, \citet{wilson75}, and \citet{sersic68} (hereafter ``King model'', ``Wilson model'' and ``S\'{e}rsic model''). \citet{mclau08} have described the three structural models in detail, here we briefly summarize some basic characteristics for them.

King model is most commonly used in studies of star clusters, which is the simple model of single-mass, isotropic, modified isothermal sphere. Wilson model is defined by an alternate modified isothermal sphere based on the ad hoc stellar distribution function of \citet{wilson75}, which has more extended envelope structures than the standard King isothermal sphere \citep{mclau08}. S\'{e}rsic model has an $R^{1/n}$ surface-density profile, which is used for parameterizing the surface brightness profiles of early-type galaxies and bulges of spiral galaxies \citep{bg11}. However, \citet{tanvir12} found that some classical GCs in M31 which exhibit cuspy core profiles are well fitted by S\'{e}rsic model of index $n\sim2-6$. The clusters with cuspy cores have usually been called post-core collapse \citep[see][and references therein]{ng06}.

\subsection{Fits}

After the intensity profiles were corrected for extinction (see Section 2.4 for details), we fitted models to the brightness profiles of the sample clusters.

We first convolved the three models with PSFs for the filters used. Given a value for the scale radius $r_0$, we compute a dimensionless model
profile $\widetilde{I}_{\rm mod}\equiv I_{\rm mod}/I_0$ and then perform the convolution,

\begin{equation}
\widetilde{I}_{\rm mod}^{*} (R | r_0) = \int\!\!\!\int_{-\infty}^{\infty}
               \widetilde{I}_{\rm mod}(R^\prime/r_0)
               \widetilde{I}_{\rm PSF}
               \left[(x-x^\prime),(y-y^\prime)\right]dx^\prime dy^\prime,
\label{eq:convol}
\end{equation}
where $R^2=x^2+y^2$, and $R^{\prime2}=x^{\prime2}+y^{\prime2}$.
$\widetilde{I}_{\rm PSF}$ was approximated using
the equation (4) of \citet{wang13} \citep[see][for details]{mclau08}. The best fitting model was derived by calculating and minimizing $\chi^2$ as the sum of squared differences between model intensities and observed intensities with extinction corrected,
\begin{equation}
\chi^2=\sum_{i}{\frac{[I_{\rm obs}(R_i)-I_0\widetilde{I}_{\rm mod}^{*}(R_i|r_0)
       -I_{\rm bkg}]^2}{\sigma_{i}^{2}}},
\end{equation}
in which a background $I_{\rm bkg}$ was also fitted. The uncertainties of observed intensities listed in Table 2 were used as weights.

We plot the fitting for the sample clusters in Figures 4--13. The observed intensity profile with extinction corrected is plotted as a function of logarithmic projected radius. The open squares are the data points included in the model fitting, while the crosses are points flagged as ``DEP'' or ``BAD'', which are not used to constrain the fit \citep{wang13}. The best-fitting models, including King model (K66), Wilson model (W), and S\'{e}rsic model (S) are shown with a red solid line from the left to the right panel, with a fitted $I_{\rm bkg}$ added. The blue dashed lines represent the shapes of the PSFs for the filters used.

\begin{figure*}[!htb]
\figurenum{4}
\center
\resizebox{\width}{!}{\rotatebox{0}
{\includegraphics[width=0.9\textwidth]{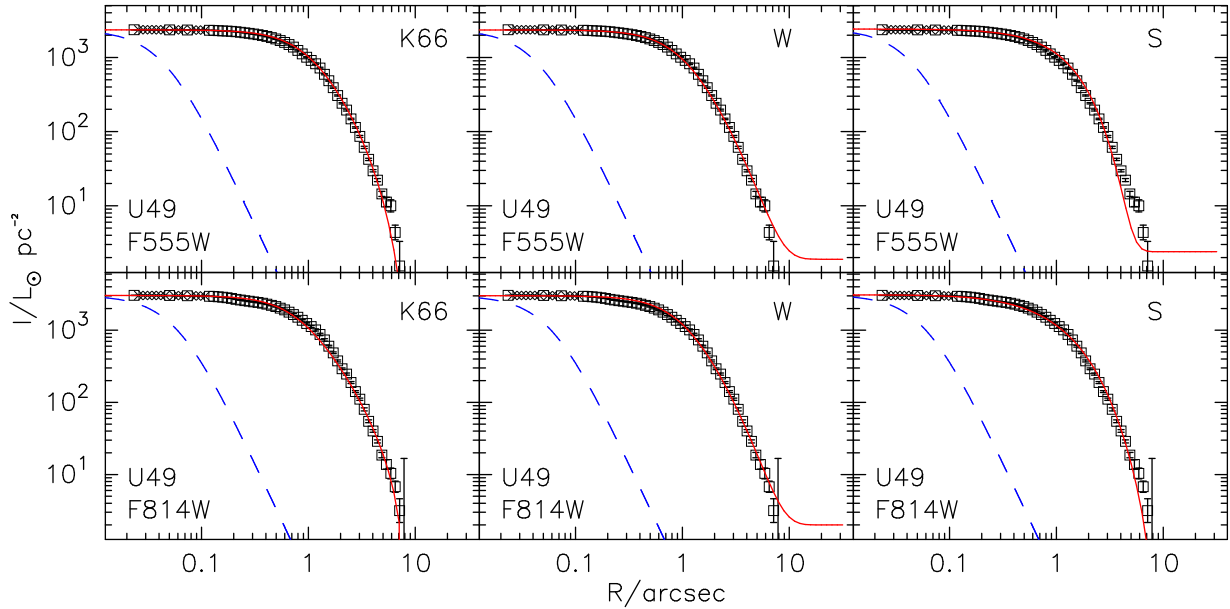}}}
\caption{Surface brightness profiles and model fits to the sample cluster U49, with the data of F555W and F814W bands from top to bottom. The three panels in each line are fits to, from left to right: King model (K66), Wilson model (W), and S\'{e}rsic model (S). The open squares are the data points included in the model fitting, while the crosses are points flagged as ``DEP'' or ``BAD'', which are not used to constrain the fit. The best-fitting models are shown with a red solid line. The blue dashed lines represent the shapes of the PSFs for the filters used.}
\label{fig4}
\end{figure*}

\begin{figure*}[!htb]
\figurenum{5}
\center
\resizebox{\width}{!}{\rotatebox{0}
{\includegraphics[width=0.9\textwidth]{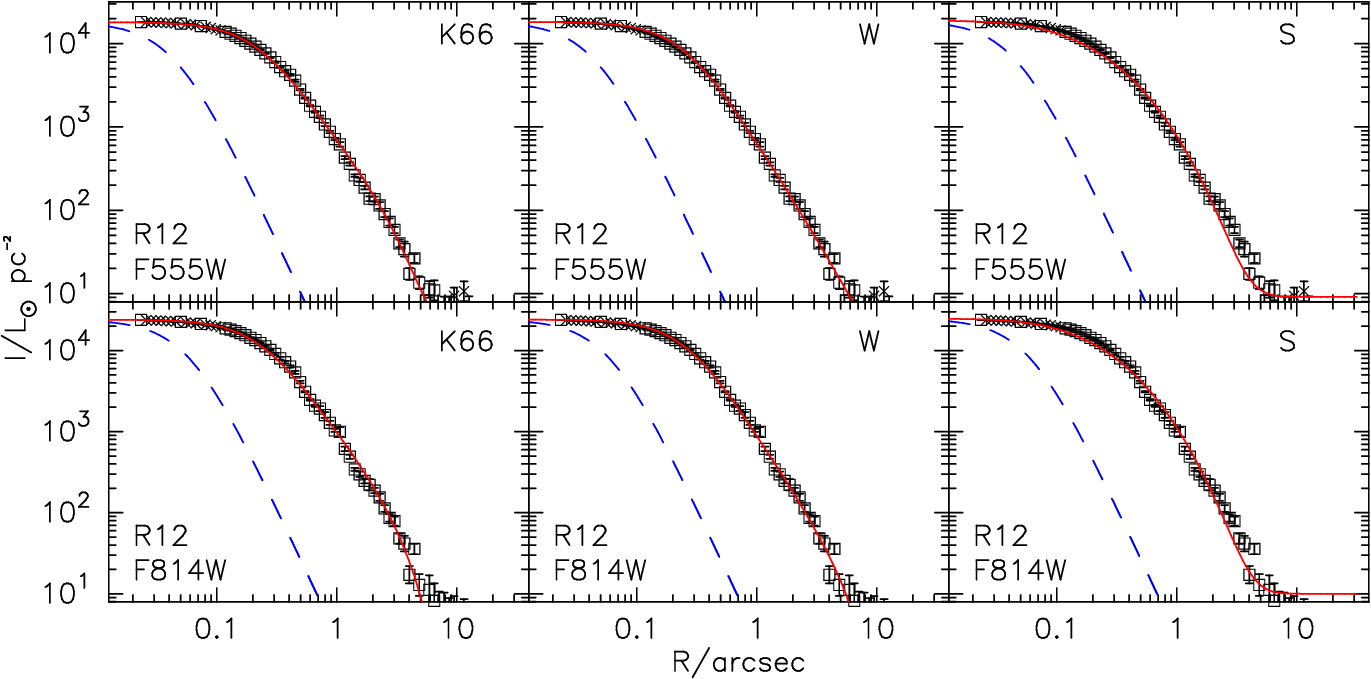}}}
\caption{Same as Fig. 4 except that surface brightness profiles and model fits to the sample cluster R12.}
\label{fig5}
\end{figure*}

\begin{figure*}[!htb]
\figurenum{6}
\center
\resizebox{\width}{!}{\rotatebox{0}
{\includegraphics[width=0.9\textwidth]{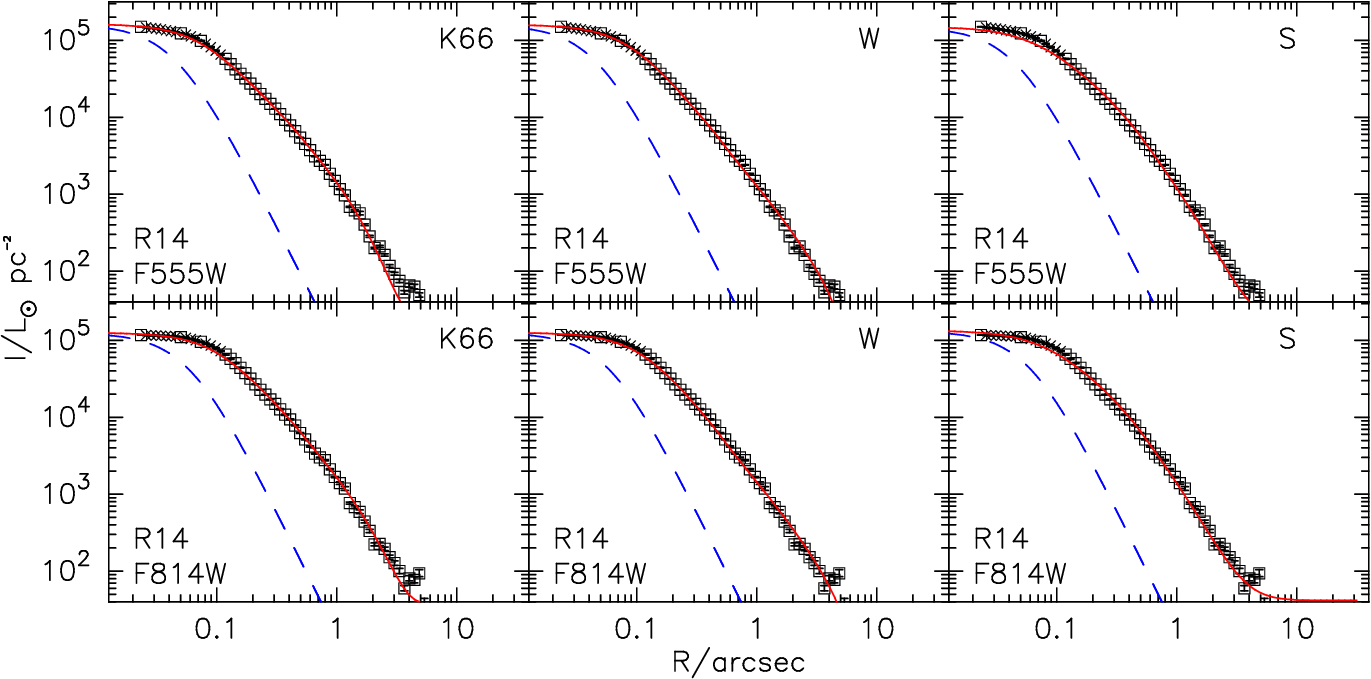}}}
\caption{Same as Fig. 4 except that surface brightness profiles and model fits to the sample cluster R14.}
\label{fig6}
\end{figure*}

\begin{figure*}[!htb]
\figurenum{7}
\center
\resizebox{\width}{!}{\rotatebox{0}
{\includegraphics[width=0.9\textwidth]{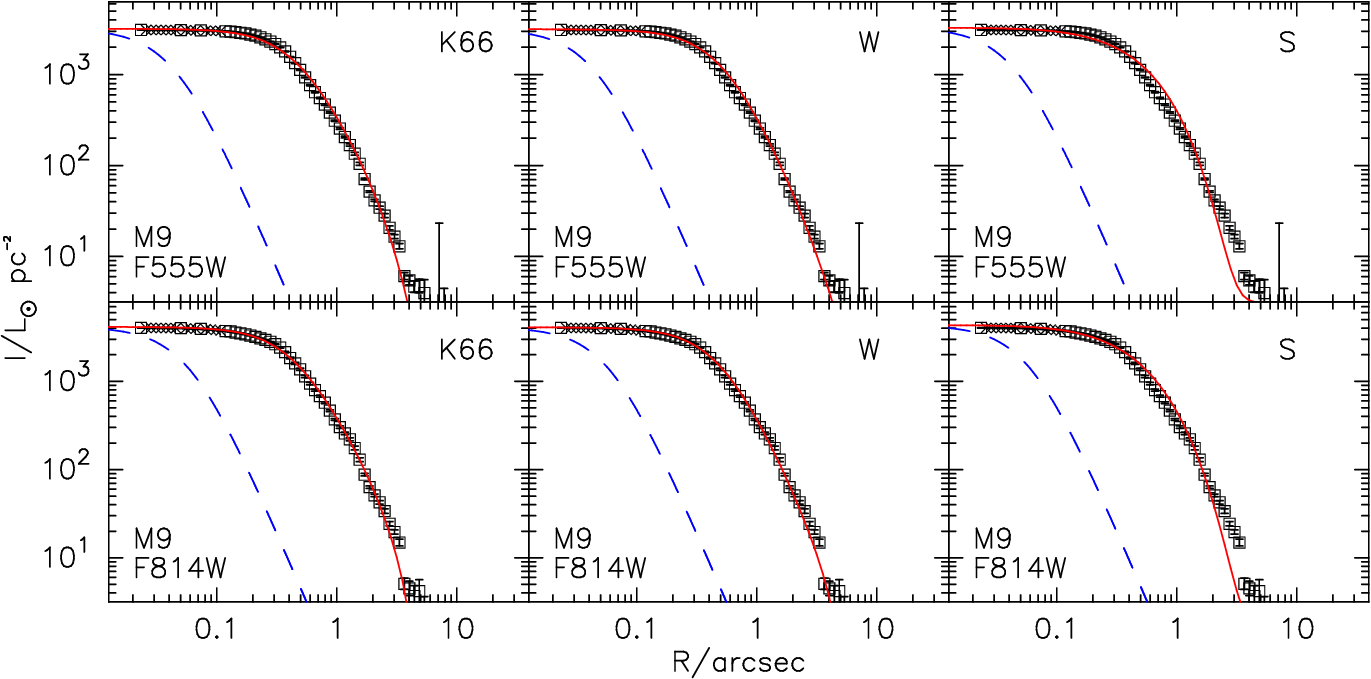}}}
\caption{Same as Fig. 4 except that surface brightness profiles and model fits to the sample cluster M9.}
\label{fig7}
\end{figure*}

\begin{figure*}[!htb]
\figurenum{8}
\center
\resizebox{\width}{!}{\rotatebox{0}
{\includegraphics[width=0.9\textwidth]{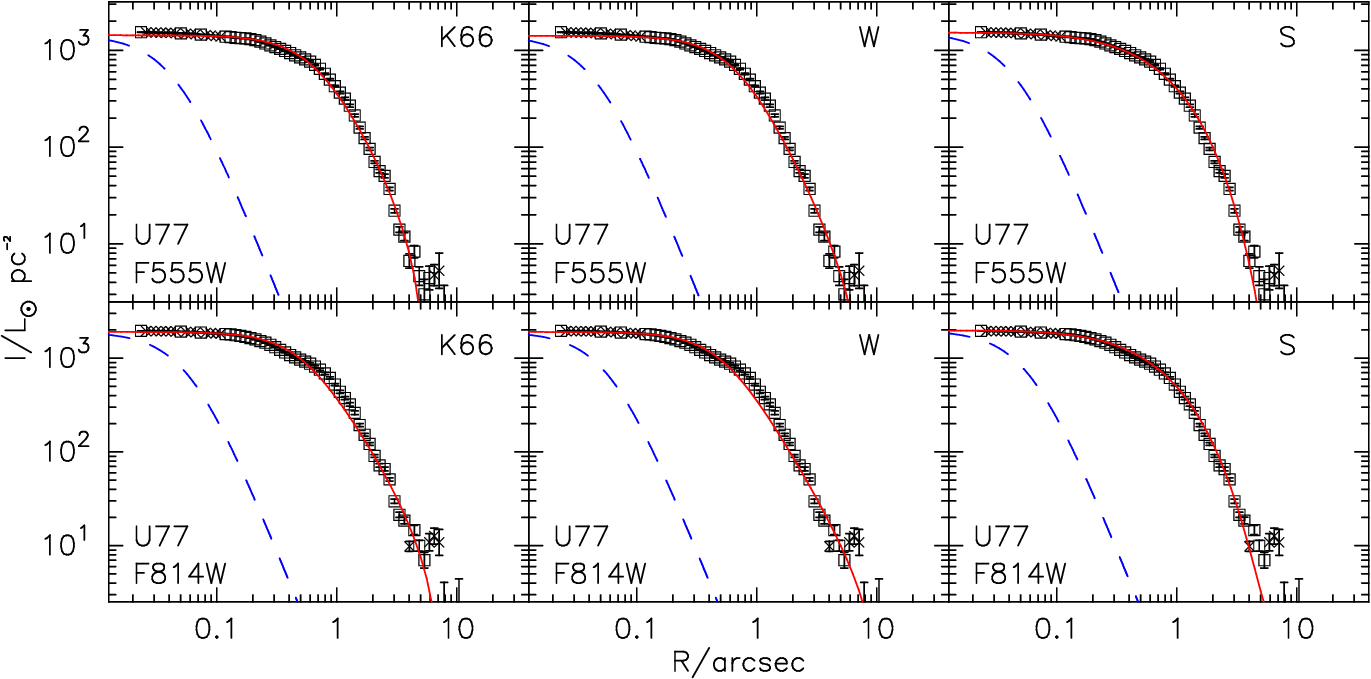}}}
\caption{Same as Fig. 4 except that surface brightness profiles and model fits to the sample cluster U77.}
\label{fig8}
\end{figure*}

\begin{figure*}[!htb]
\figurenum{9}
\center
\resizebox{\width}{!}{\rotatebox{0}
{\includegraphics[width=0.9\textwidth]{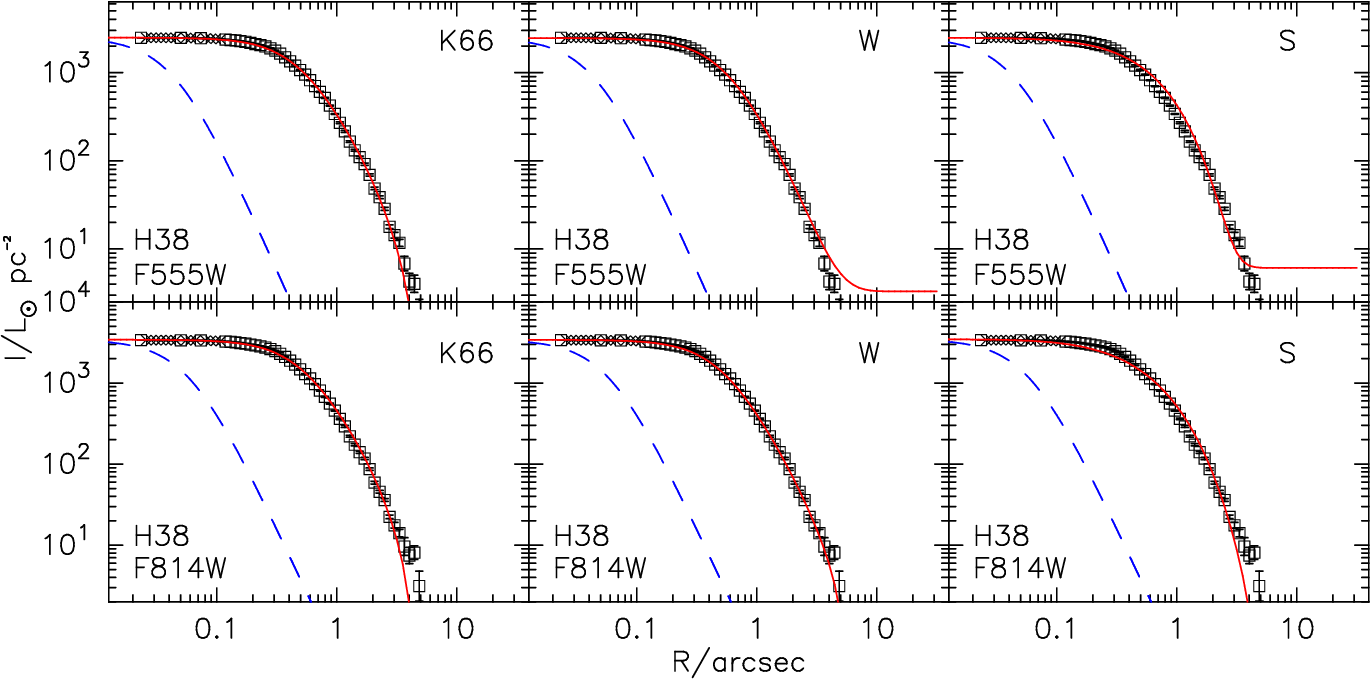}}}
\caption{Same as Fig. 4 except that surface brightness profiles and model fits to the sample cluster H38.}
\label{fig9}
\end{figure*}

\begin{figure*}[!htb]
\figurenum{10}
\center
\resizebox{\width}{!}{\rotatebox{0}
{\includegraphics[width=0.9\textwidth]{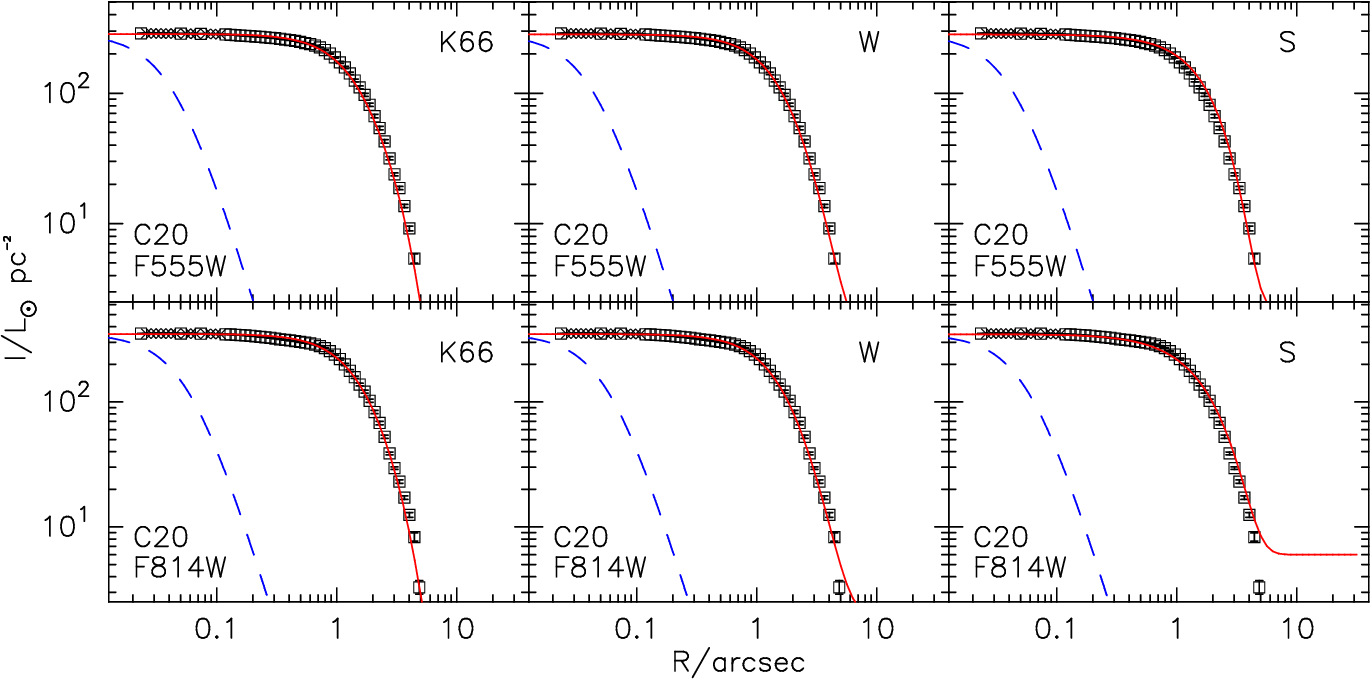}}}
\caption{Same as Fig. 4 except that surface brightness profiles and model fits to the sample cluster C20.}
\label{fig10}
\end{figure*}

\begin{figure*}[!htb]
\figurenum{11}
\center
\resizebox{\width}{!}{\rotatebox{0}
{\includegraphics[width=0.9\textwidth]{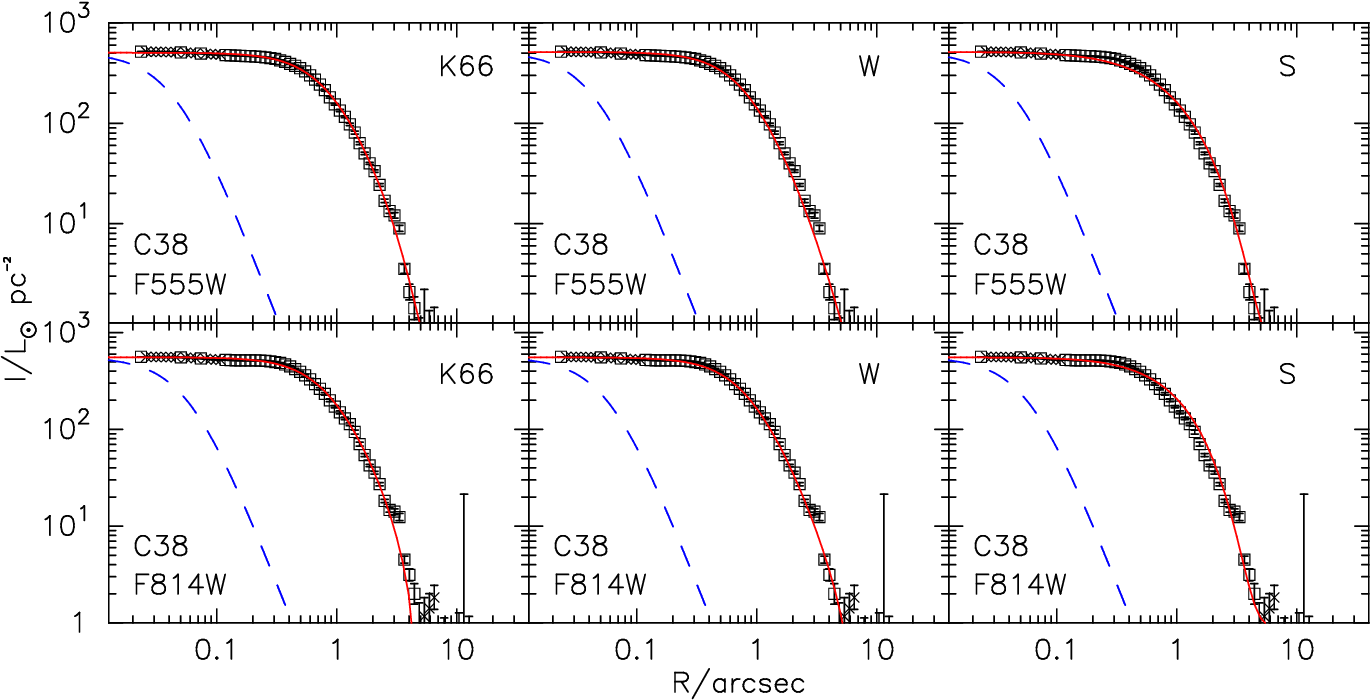}}}
\caption{Same as Fig. 4 except that surface brightness profiles and model fits to the sample cluster C38.}
\label{fig11}
\end{figure*}

\begin{figure*}[!htb]
\figurenum{12}
\center
\resizebox{\width}{!}{\rotatebox{0}
{\includegraphics[width=0.9\textwidth]{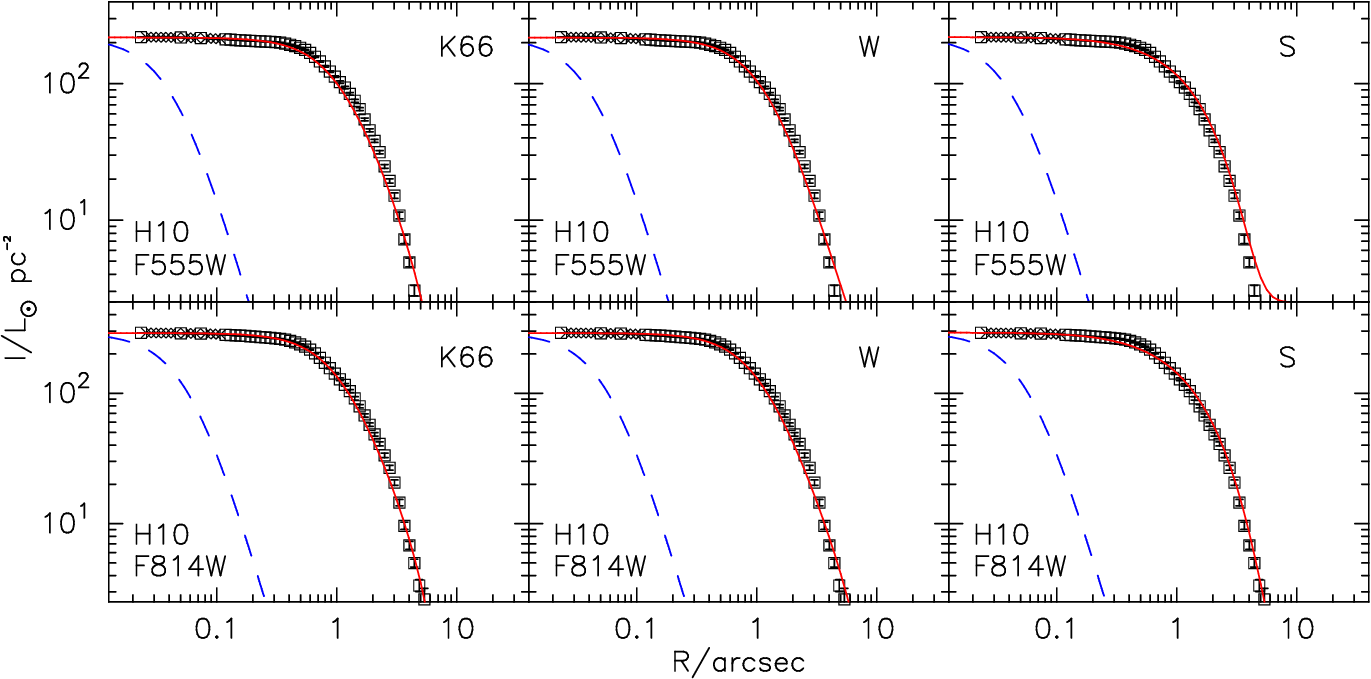}}}
\caption{Same as Fig. 4 except that surface brightness profiles and model fits to the sample cluster H10.}
\label{fig12}
\end{figure*}

\begin{figure*}[!htb]
\figurenum{13}
\center
\resizebox{\width}{!}{\rotatebox{0}
{\includegraphics[width=0.9\textwidth]{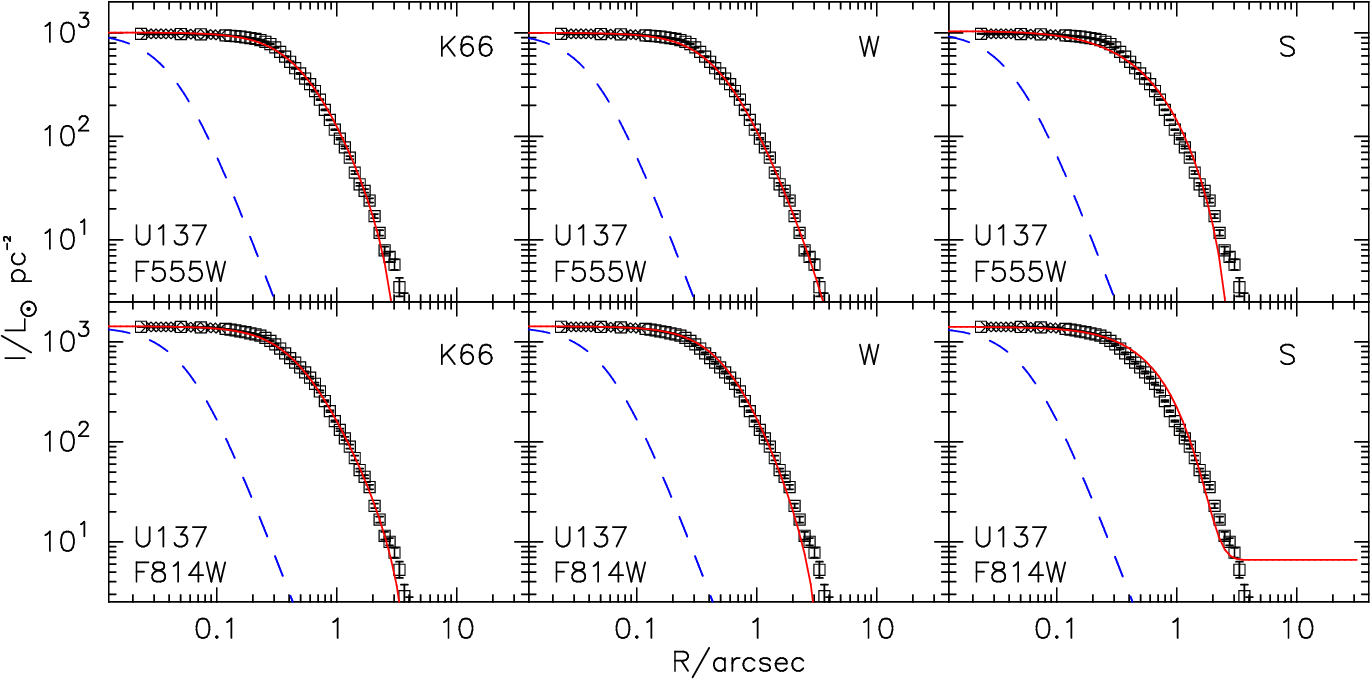}}}
\caption{Same as Fig. 4 except that surface brightness profiles and model fits to the sample cluster U137.}
\label{fig13}
\end{figure*}

As \citet{wang13} did, the values of mass-to-light ratios ($M/L$ values), which were used to derive the dynamical parameters, were determined from the population-synthesis models of \citet{bc03}, assuming a \citet{chab03} initial mass function. The ages and metallicities used to compute $M/L$ values in the $V$ band are listed in Table 1.

The basic structural and various derived dynamical parameters of the best-fitting models for each cluster are listed in Tables 3--5, with a description of each parameter/column at the end of each table \citep[see][for details of the calculation]{mclau08}. The uncertainties of these parameters were estimated by calculating their variations in each model that yields $\chi^2$ within 1 of the global minimum for a cluster, while combined in quadrature with the uncertainties in $M/L$ for the parameters related to it \citep[see][for details]{mm05}.

At last, we constructed and listed the fundamental plane of the sample clusters in Table 6, defined by \citet{mclau08} using mass surface density instead of luminosity surface density in the ``$\kappa$'' parameters defined by
\citet{bender92}.
\begin{equation}
\kappa_{m,1}=(\log \sigma_{p,0}^2 + \log R_h)/\sqrt{2}
\end{equation}
\begin{equation}
\kappa_{m,2}=(\log \sigma_{p,0}^2 + \log \Sigma_h - \log R_h)/\sqrt{6}
\end{equation}
\begin{equation}
\kappa_{m,3}=(\log \sigma_{p,0}^2 + \log \Sigma_h - \log R_h)/\sqrt{3}
\end{equation}

\subsection{Comparison to Previous Results}

\citet{larsen02} presented ellipticities, position angles and surface brightness profiles for four sample clusters (H38, M9, R12 and U49), and they also determined structural parameters using \citet{king66} model fits for them. These authors used the same $HST$ images as we used here. They filtered the images of the clusters with an $11\times11$ $\rm{pixel^2}$ median filter. \citet{larsen02} presented the ellipticities to be less than $\sim 0.1$ for H38, M9 and R12 comparing with $\sim 0.1$ for M9 and R12 and $\sim 0.2$ for H38 obtained here. \citet{larsen02} presented the ellipticity of U49 to be $\epsilon=0.10-0.15$ comparing with $\sim 0.25$ obtained here. In fact, \citet{larsen02} only plotted the distributions of ellipticities and position angles for four clusters between 0.45 and 4.2 arcsec, and did not calculate the average values. When we plot the distributions of ellipticities and position angles for them between 0.45 and 4.2 arcsec, we find that, except for R12, the distributions of ellipticities and position angles for the other three sample clusters obtained here are in agreement with those of \citet{larsen02}. We indicated in Section 2.2 that we did not filter the images of R12 because its surface brightness profiles are smooth enough (see Figure 5). However, beyond $\sim 1''$ in the F555W band and $\sim 2''$ in the F814W band, the ellipticity and the position angle of R12 cannot be derived here because of low signal-to-noise ratio. Figure 14 compares the values of cluster parameters obtained here with those in \citet{larsen02}, including the concentration $c$, scale radius $r_0$, and central surface brightness $\mu_{0}$. It is evident that, in general, our results are in agreement with those derived by \citet{larsen02}, and systematic differences are not found. In addition, \citet{CBF99a,CBF99b} derived the core radius of U137 (their No. 54) in the F555W band. \citet{roman12} derived the structural parameters including the core radius, concentration, half-light radius and central surface brightness for U137 (their No. 22). In general, our results are in agreement with those in previous studies. The offset of the core radius of U137 between this paper and \citet{CBF99a,CBF99b} (this study minus previous study) is 0.43 pc in the F555W band. \citet{roman12} derived the core radius, concentration, half-light radius and central surface brightness of U137 in the F606W and F814W bands. The offsets of the core radius, concentration, half-light radius and central surface brightness of U137 between this paper and \citet{roman12} (this study minus previous study) are $-0.33$ pc, 0.11, $-0.50$ pc, and 0.14 mag in the F555W (F606W) band, and $-0.64$ pc, 0.26, $-0.42$ pc, and $-0.22$ mag in the F814W band. The cluster structural parameters used to compare here were obtained in \citet{larsen02}, \citet{roman12},  and this paper by fitting King model. \citet{roman12} obtained the ellipticity of U137 to be 0.06 in the F606W band comparing with 0.10 in the F555W band obtained here, and to be 0.09 in the F814W band comparing with 0.13 obtained here. However, when we calculate the final ellipticity of U137 as the average of the values between 0.5 and 1.5 arcsec as \citet{roman12} did, we derive the ellipticity of U137 to be 0.05 and 0.11 in the F555W and F814 bands, respectively, which are in good agreement with those of \citet{roman12}.

\begin{figure*}[!htb]
\figurenum{14}
\center
\resizebox{\width}{!}{\rotatebox{-90}
{\includegraphics[width=0.7\textwidth]{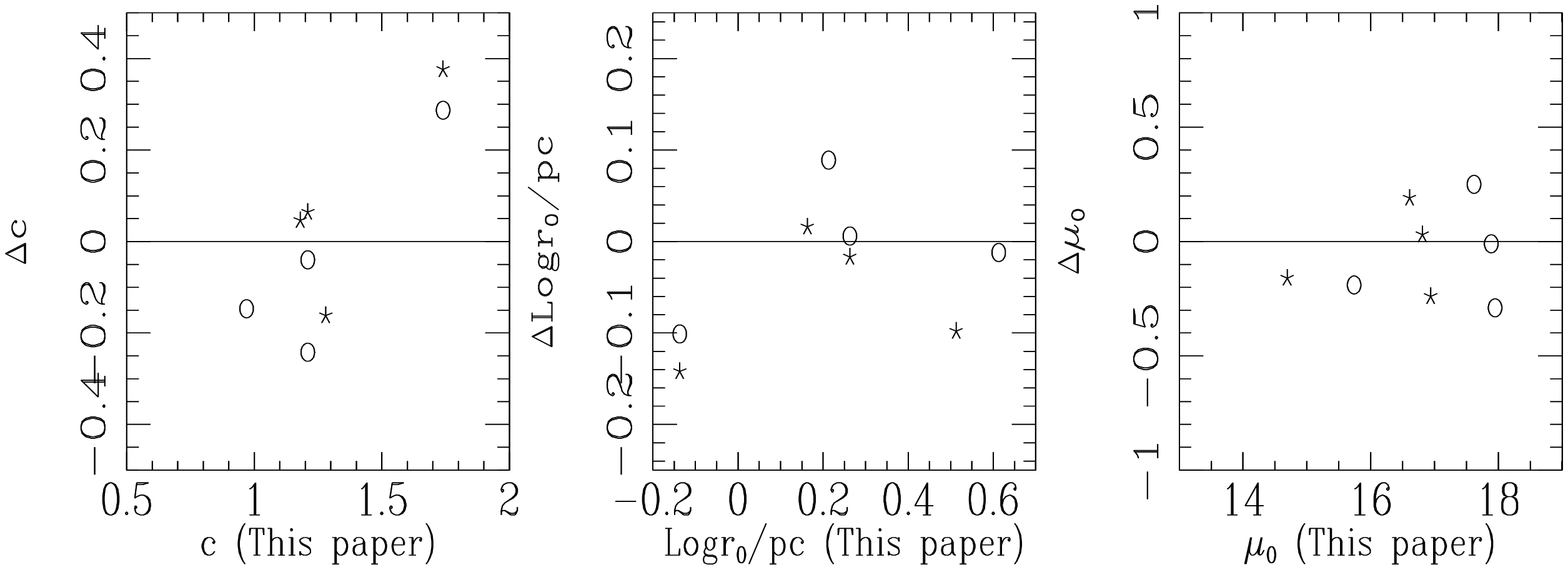}}}
\vspace{-4.5cm}
\caption{Comparison of our newly obtained cluster structural parameters by fitting King model \citep{king66} with previous measurements by \citet{larsen02}. From left to right: concentration, scale radius, and central surface brightness. Open circles: in the F555W band; stars: in the F814W band.}
\label{fig14}
\end{figure*}

\subsection{Predicted Velocity Dispersion}

One eventual use for the material presented in Section 3.2 should be to facilitate the comparison of dynamical star cluster $M/L$ values to the population-synthesis model values that we calculated here. For the star clusters of M33, the determinations of $M/L$ values are very limited: only \citet{larsen02} measured observed velocity dispersions for four sample clusters using high-dispersion spectra from the HIRES echelle spectrograph on the Keck I telescope. These authors determined the $M/L_V$ values for these four clusters. The upper panel of Figure 15 shows the observed versus model velocity dispersions for the sample clusters in question. The lower panel then shows the square of the ratio of observed to predicted dispersions, which is equal to the ratio of dynamical to population-synthesis $M/L$. Figure 15 shows that there is generally good agreement between the observed and predicted velocity dispersions except for R12. However, it is true that, because of the limited sample clusters, this conclusion should be confirmed based on a large sample objects. The mean ratio between the observed and predicted velocity dispersions is $\sigma_{\rm obs}/\sigma_{\rm pred}=0.91$, with an interquartile range of $\pm0.08$. This corresponds to a ratio between dynamical and population-synthesis-derived mass-to-light ratios of $\Upsilon^{\rm dyn}_{V}/\Upsilon^{\rm pop}_{V}=0.84\pm0.15$, consistent with the values for this ratio of $0.82\pm0.07$ presented by \citet{mm05} for Milky Way and old Magellanic Cloud clusters, $0.73\pm0.25$ presented by \citet{barmby07} for M31 GCs, and $0.81\pm0.40$ published by \citet{mclau08} for NGC 5128 GCs.

\begin{figure*}[!htb]
\figurenum{15}
\center
\resizebox{\width}{!}{\rotatebox{-90}
{\includegraphics[width=0.7\textwidth]{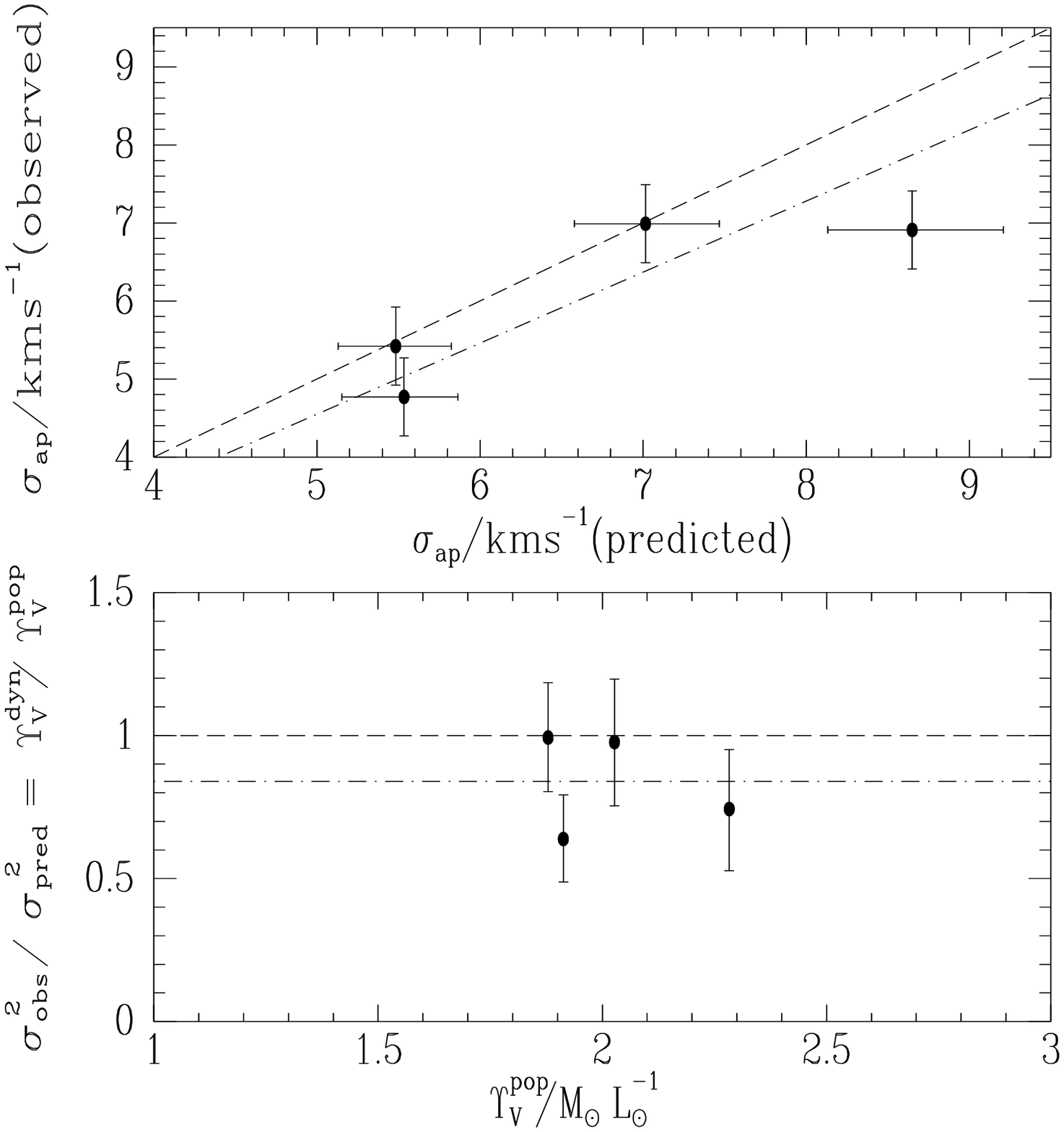}}}
\vspace{0.0cm}
\caption{{\it Top}: Comparison of structural-model predictions for central velocity dispersion with observations from \citet{larsen02}. The dashed line is the line of equality; the dot-dashed line is the mean ratio $\sigma_{\rm obs}/\sigma_{\rm pred}=0.91$. {\it Bottom}: Comparison of dynamical and population-synthesis-derived mass-to-light ratios. The dashed line is the line of equality; the dot-dashed line is the mean ratio $\Upsilon^{\rm dyn}_{V}/\Upsilon^{\rm pop}_{V}=0.84$.}
\label{fig15}
\end{figure*}

\begin{figure*}[!htb]
\figurenum{16}
\center
\resizebox{\width}{!}{\rotatebox{-90}
{\includegraphics[width=0.7\textwidth]{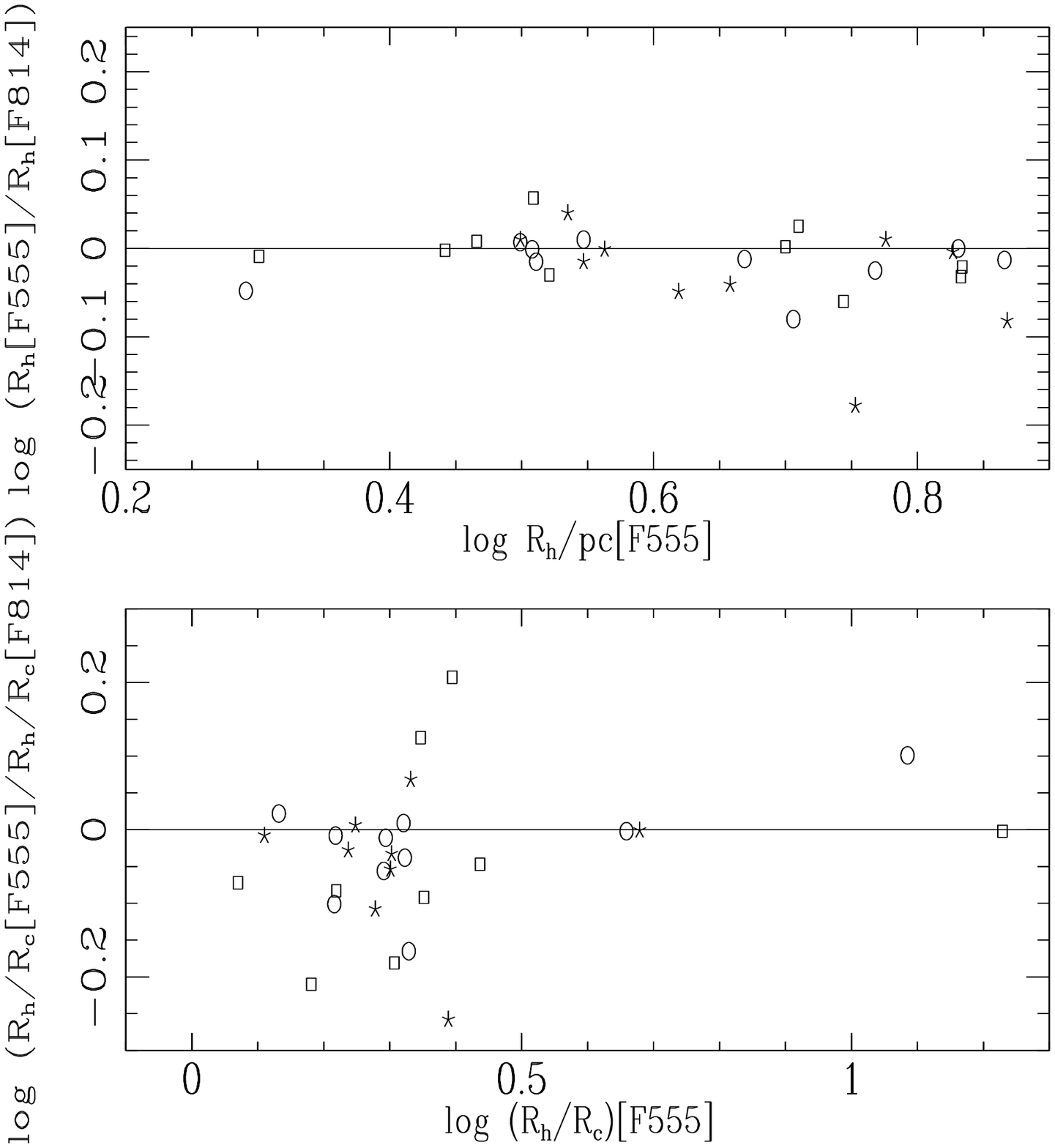}}}
\vspace{0.0cm}
\caption{Comparison of structural parameters for model fits to the sample clusters observed in both F555W and F814W bands. From top to bottom: projected half-light radius and ratio of half-light to core radius. Open circles: \citet{king66} model; squares: \citet{sersic68} model; stars \citet{wilson75} model.}
\label{fig16}
\end{figure*}

\subsection{Comparison of Fits in the F555W and F814W Bands}

Comparing model fits to the same cluster observed in different bands allows assessment of the systematic errors and color dependencies in the fits \citep{barmby07}. Figure 16 compares the parameters derived from fits to the sample clusters observed in both F555W and F814W bands. In general, the agreement is quite good, with somewhat larger scatter for the \citet{sersic68} and \citet{wilson75} model fits than for the \citet{king66} model fits. This result is in agreement with those presented by \citet{barmby07} for M31 star clusters. However, for some sample clusters, both projected half-light radius and ratio of half-light to core radius as fit to the F555W data are a little smaller than those from the F814W data for the three models.


Because the model-fit results in the two WFPC2 bands are quite similar, we only consider the F555W model fits from now on. In addition, fits to clusters in the Milky Way are performed in the $V$ band, allowing us direct comparison without being concerned about possible color gradients (See Section 4).

\begin{figure*}[!htb]
\figurenum{17}
\center
\resizebox{\width}{!}{\rotatebox{0}
{\includegraphics[width=0.9\textwidth]{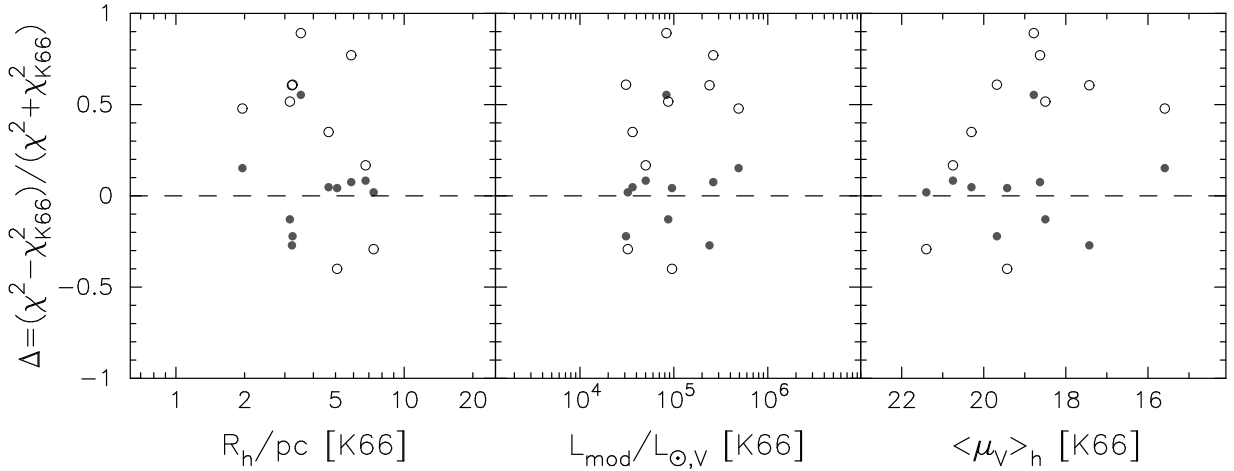}}}
\caption{Relative quality of fit for Wilson and S\'{e}rsic models (filled and open circles) versus King model for the sample clusters in this paper.}
\label{fig17}
\end{figure*}

\subsection{Comparison of Three Model Fittings}

In order to determine which model can fit the structure of clusters best, \citet{mm05} and \citet{mclau08} defined a relative $\chi^2$ index, $\Delta$, which compares the $\chi^2$ of the best fit of an ``alternate'' model with the $\chi^2$ of the best fit of King model,
\begin{equation}
\Delta=(\chi^2_{\rm alternate}-\chi^2_{\rm K66})/
(\chi^2_{\rm alternate}+\chi^2_{\rm K66}).
\end{equation}

If the parameter $\Delta$ is zero, the two models fit the same cluster equally well. Positive values indicate a better fit of King model, and negative values indicate the ``alternate'' model is a better fit than King model. Figure 17 shows the relative quality of fit, $\Delta$ for Wilson- and S\'{e}rsic-model fits (filled and open circles, respectively, in all panels) versus King-model fits for the sample clusters in this paper. The $\Delta$ values are plotted as a function of some structural parameters, including the half-light radius $R_h$, total model luminosity $L_{\rm mod}$, and the intrinsic average surface brightness $\langle\mu_V\rangle_h=26.358-2.5\log(L_V/2\pi R^2_h)$. In general, \citet{king66} model fits the M33 halo cluster data better than \citet{sersic68} model. In addition, there is only one cluster which has $\Delta > 0.5$ for Wilson-model fits. The remaining clusters have absolute values of $\Delta < 0.2$ for Wilson-model fits. So, we consider that King and Wilson models fit equally well for nearly all of the sample clusters in M33 (except for one). This result differs from those for the clusters in NGC 5128 \citep{mclau08} and M31 \citep{barmby07,wang13}. \citet{mclau08} indicated that for the clusters in NGC 5128, \citet{wilson75} model fits as well as or better than \citet{king66} model, however, \citet{barmby07} and \citet{wang13} showed that for most clusters in M31, \citet{king66} model fits better than \citet{wilson75} model. In addition, Figure 17 immediately shows that \citet{wilson75} model fits the M33 sample cluster data better than \citet{sersic68} model. Based on our small sample clusters of M33, we cannot see these structural parameters strongly affect $\Delta$. For M31 and NGC 5128 clusters, the results are somewhat different. For M31 clusters, \citet{barmby07} showed that the bright clusters ($L_{\rm mod}>10^5L_{\odot}$) are better fitted by \citet{king66} model, however, for NGC 5128 clusters, \citet{mclau08} showed that the bright clusters are generally fitted much better by \citet{wilson75} and \citet{sersic68} models. In addition, \citet{mm05} presented a database of structural and dynamical properties for 153 spatially resolved star clusters in the Milky Way, the Large and Small Magellanic Clouds, and the Fornax dwarf spheroidal. The results of \citet{mm05} showed that globulars and young massive clusters in theses galaxies are systematically better fitted by the extended \citet{wilson75} model than by \citet{king66} model. \citet{sersic68} model often fits about as well as Wilson model but can be significantly worse. \citet{roman12} derived structural parameters for 161 star clusters using the {\it HST} images. These authors found that young clusters (${\rm log~age < 8}$) are notably better fitted by Elson-Freeman-Fall model \citep{elson87} with no radial truncation, while older clusters show no significant differences between \citet{king62} and \citet{elson87} fits.

\begin{figure*}[!htb]
\figurenum{18}
\center
\resizebox{\width}{!}{\rotatebox{0}
{\includegraphics[width=0.85\textwidth]{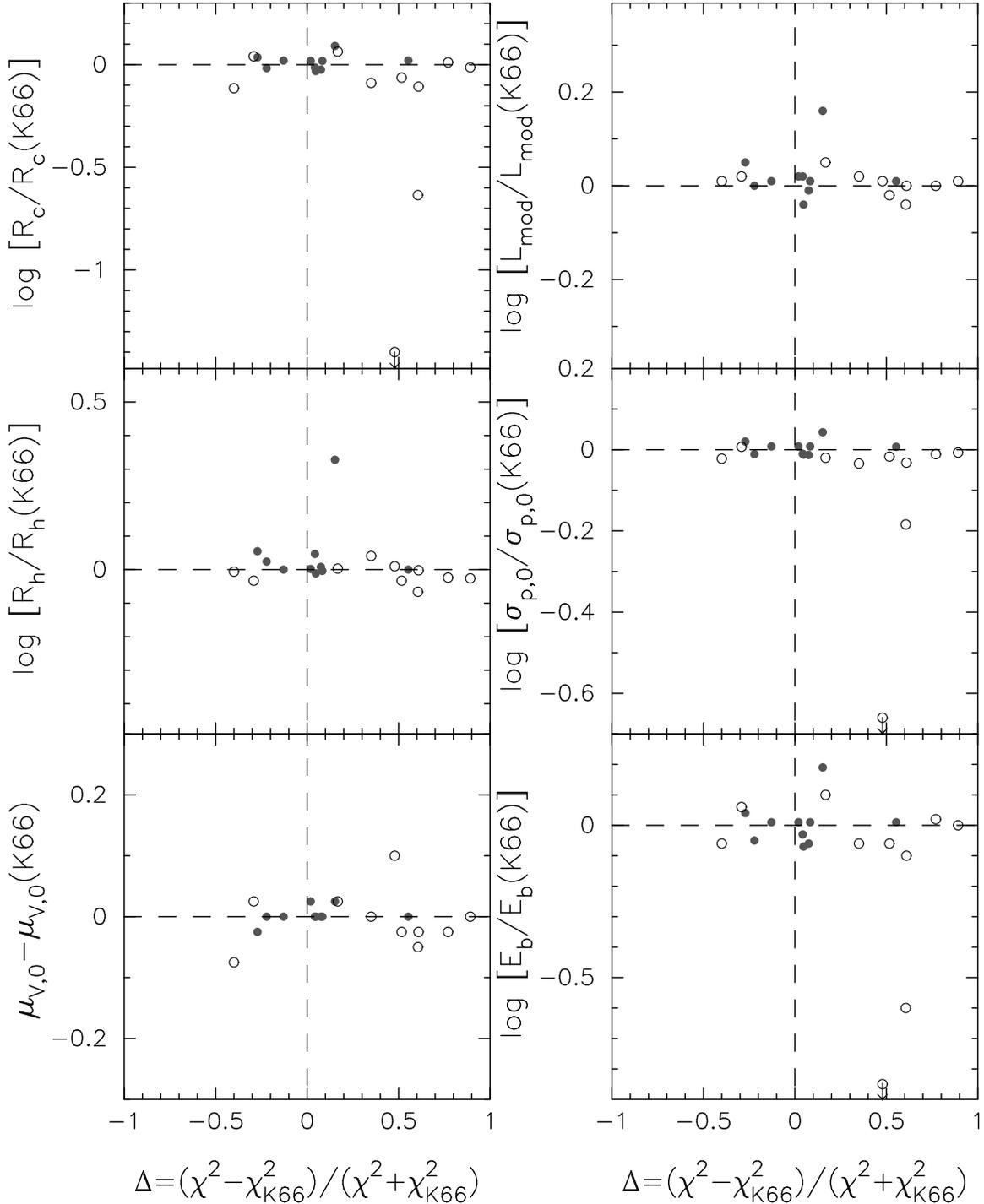}}}
\caption{Comparison of some structural and dynamical parameters for Wilson and S\'{e}rsic models versus King model for the sample clusters here, including the projected core radius $R_c$, projected half-light radius $R_h$, central surface brightness $\mu_{V,0}$ in the $V$ band, total model luminosity $L_{\rm mod}$, predicted central line-of-sight velocity dispersion $\sigma_{p,0}$, and the global binding energy $E_b$. Symbols are as in Fig. 17.}
\label{fig18}
\end{figure*}

Figure 18 compares the relative quality of fit, $\Delta$ values with a number of structural parameters ($R_c$, $R_h$, $\mu_{V,0}$, $L_{\rm mod}$, $\sigma_{p,0}$, and $E_b$) for the sample clusters here. It is evident that the parameters do not clearly vary with goodness of fit. There is one cluster (R14) with comparable $\chi^2$, but large discrepancy of $R_c$, $\sigma_{p,0}$, and $E_b$ values for King- and S\'{e}rsic-model fits. Arrows are attached to the points for R14 in the figure when the parameter differences fall outside the plotted range in any panel. In addition, most parameters derived from Wilson model are slightly larger than those from King model, while parameters derived from S\'{e}rsic model are smaller than those from King model, especially for $R_c$, $\sigma_{p,0}$, and $E_b$. The mean offsets between parameters derived for the same cluster from \citet{king66} and \citet{wilson75} models are: $\delta(\log R_c)=0.012\pm0.011$, $\delta(\log R_h)=0.045\pm0.032$, $\delta\mu_{V,0}=0.002\pm0.004$, $\delta(\log L_V)=0.023\pm0.017$, $\delta(\log\sigma_{p,0})=0.005\pm0.03$, and $\delta(\log E_b)=0.006\pm0.023$. The mean offsets from \citet{king66} and \citet{sersic68} models are: $\delta(\log R_c)=-0.101\pm0.070$ (R14 aside), $\delta(\log R_h)=-0.014\pm0.009$, $\delta\mu_{V,0}=-0.005\pm0.015$, $\delta(\log L_V)=0.006\pm0.008$, $\delta(\log\sigma_{p,0})=-0.036\pm0.026$ (R14 aside), and $\delta(\log E_b)=-0.078\pm0.069$ (R14 aside).

\section{DISCUSSION}
\label{discussion.sec}

We combined the newly derived structural and dynamical parameters of M33 halo GCs by King model fits here with those derived by King model fits for M31 GCs \citep{barmby07,wang13}, and Milky Way GCs \citep{mm05} to construct a large sample to look into the correlations between the parameters. The ellipticities and galactocentric distances for Milky Way GCs are from \citet{harris96} (2010 edition). The parameters used in the following discussion for M31 GCs \citep{barmby07,wang13} are those derived on the band close to the $V$ band (e.g., F555W and F606W).

\begin{figure*}[!htb]
\figurenum{19}
\center
\resizebox{\width}{!}{\rotatebox{0}
{\includegraphics[width=0.9\textwidth]{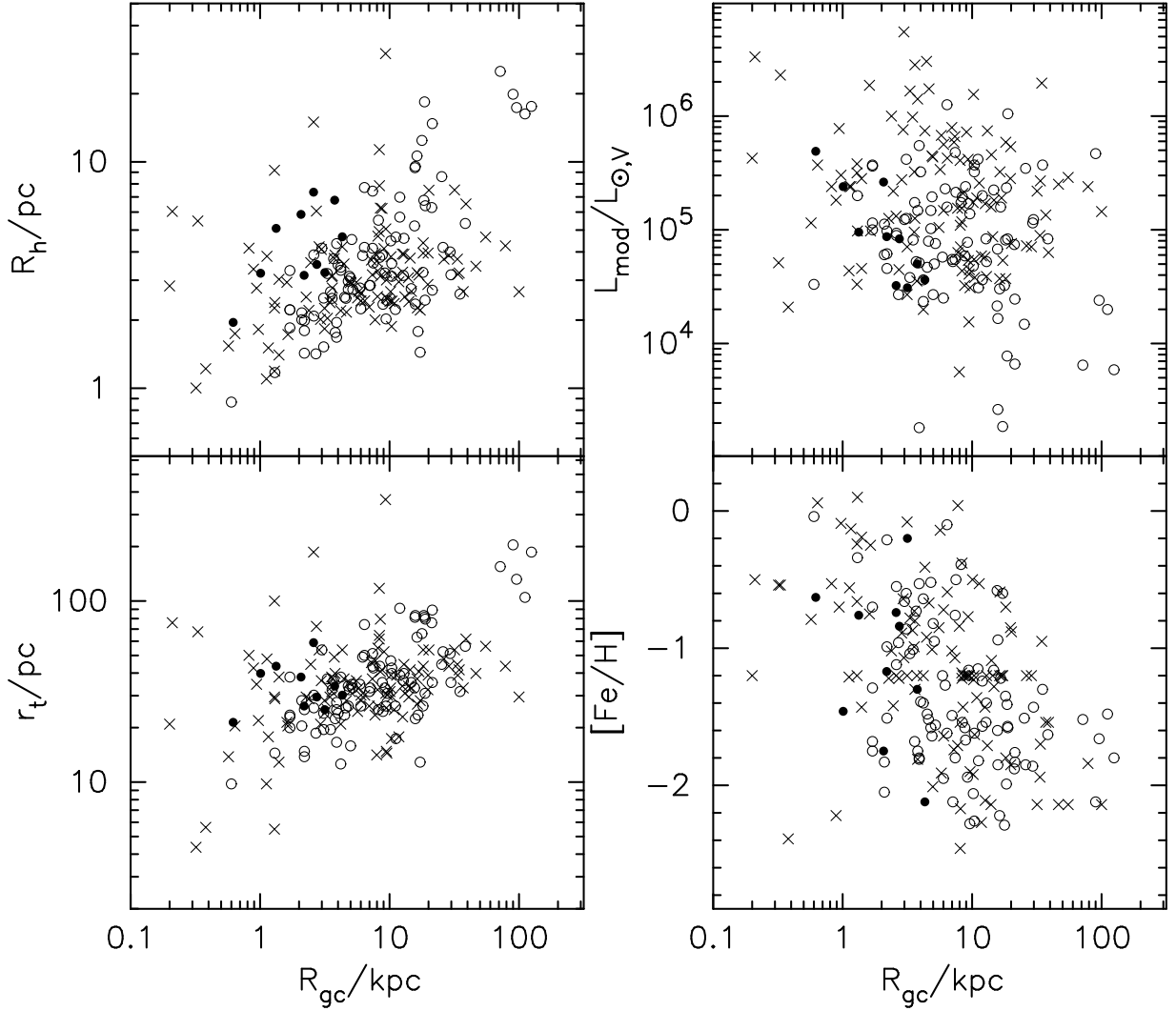}}}
\caption{Structural parameters vs. galactocentric distance $R_{\rm gc}$. The filled circles are the sample clusters in M33, the open circles are Galactic GCs \citep{mm05}, the crosses are M31 GCs \citep{barmby07,wang13}.}
\label{fig19}
\end{figure*}

\subsection{Galactocentric Distance}

Figure 19 shows structural parameters as a function of galactocentric distance $R_{\rm gc}$ for the new large sample clusters in M33, M31 and the Milky Way. For Milky Way clusters the true three-dimensional distance is used, while for M31 and M33 clusters only projected distances are available. It is evident that all the clusters in the three galaxies present similar trends, although M33 sample clusters are limited in smaller galactocentric distances. However, M33 clusters have slightly larger $R_h$ and $R_t$ than M31 and Milky Way clusters for a given $R_{\rm gc}$, although all the clusters in three galaxies follow the general trend. On the contrary, M33 clusters have slightly smaller $L_{\rm mod}$ than M31 and Milky Way clusters for a given $R_{\rm gc}$. We fitted the data of the Milky Way and M31 and of M33 to linear relations, respectively. And the results showed that there exit some offsets between the different trends. The fitting results are as follow. $R_h=0.323 \times R_{\rm gc}+0.266$ and $R_h=0.495 \times R_{\rm gc}+0.398$ for Milky Way and M31 clusters and for M33 clusters, respectively. $r_t=1.331 \times R_{\rm gc}+0.254$ and $r_t=1.505 \times R_{\rm gc}+0.055$ for Milky Way and M31 clusters and for M33 clusters, respectively. $L_{\rm mod}=5.213 \times R_{\rm gc}-0.218$ and $L_{\rm mod}=5.377 \times R_{\rm gc}-1.332$ for Milky Way and M31 clusters and for M33 clusters, respectively. We argued that this difference results from small number statistic of M33 clusters. In addition, some global trends can be seen. Clusters at larger galactocentric distances, on average, have larger $R_h$ and $r_t$ than do those which are located closer to the galactic center. The correlation between $R_h$ and $R_{\rm gc}$ showed in Figure 19 is a general property of GC systems \citep[see][and references therein]{blom12}, which results from physical conditions at the time of cluster formation as suggested by \citet{vanden91} for Milky Way GCs. \citet{harris09} considered that the scale sizes of GCs appear to be determined jointly and universally by their metallicity, spatial locations, and their mass, which are included in their conditions of formation. In fact, it is true that star clusters with small $R_{\rm gc}$ have small $R_h$ due to tidal disruption. It should be mentioned, however, that no clear correlation between $R_h$ and $R_{\rm gc}$ exists for the GCs in NGC 4649 \citep{str12}. So, \citet{str12} suggested that the sizes of GCs are not generically set by tidal limitation. The luminosities of clusters decrease with increasing $R_{\rm gc}$, implying that either strong dynamical friction drives predominantly more massive clusters inwards, or more massive clusters have a bias in favor of forming in the nuclear regions of galaxies where the higher ambient pressure and density favor the formation of more massive star clusters \citep[see][and references therein]{Georgiev09}. The lack of faint clusters at small distances from the galactic centers may be due to selection effects, since faint clusters are difficult to see against the bright background near galactic centers \citep{barmby07}. The metallicities of clusters decrease with increasing $R_{\rm gc}$, which indicates that metal-rich clusters are typically located at smaller galactocentric radii than metal-poor ones, although with large scatter. The presence or absence of a radial trend in the metallicity of a GC sample in a galaxy is an important test of galaxy formation theories \citep{bh00}. In fact, some works have presented the metallicity of GCs in the Milky Way, M31 and M33 as a function of galactocentric radius. \citet{taft89} confirmed the presence of a modest {\bf metallicity} gradient within the system of disk clusters in the Milky Way. There are some inconsistent conclusions for M31 GCs \citep[see][and references therein]{fan08}. \citet{Ma04a} did not find the presence of a modest metallicity gradient for old star clusters in M33.

\begin{figure*}[!htb]
\figurenum{20}
\center
\resizebox{\width}{!}{\rotatebox{-90}
{\includegraphics[width=0.75\textwidth]{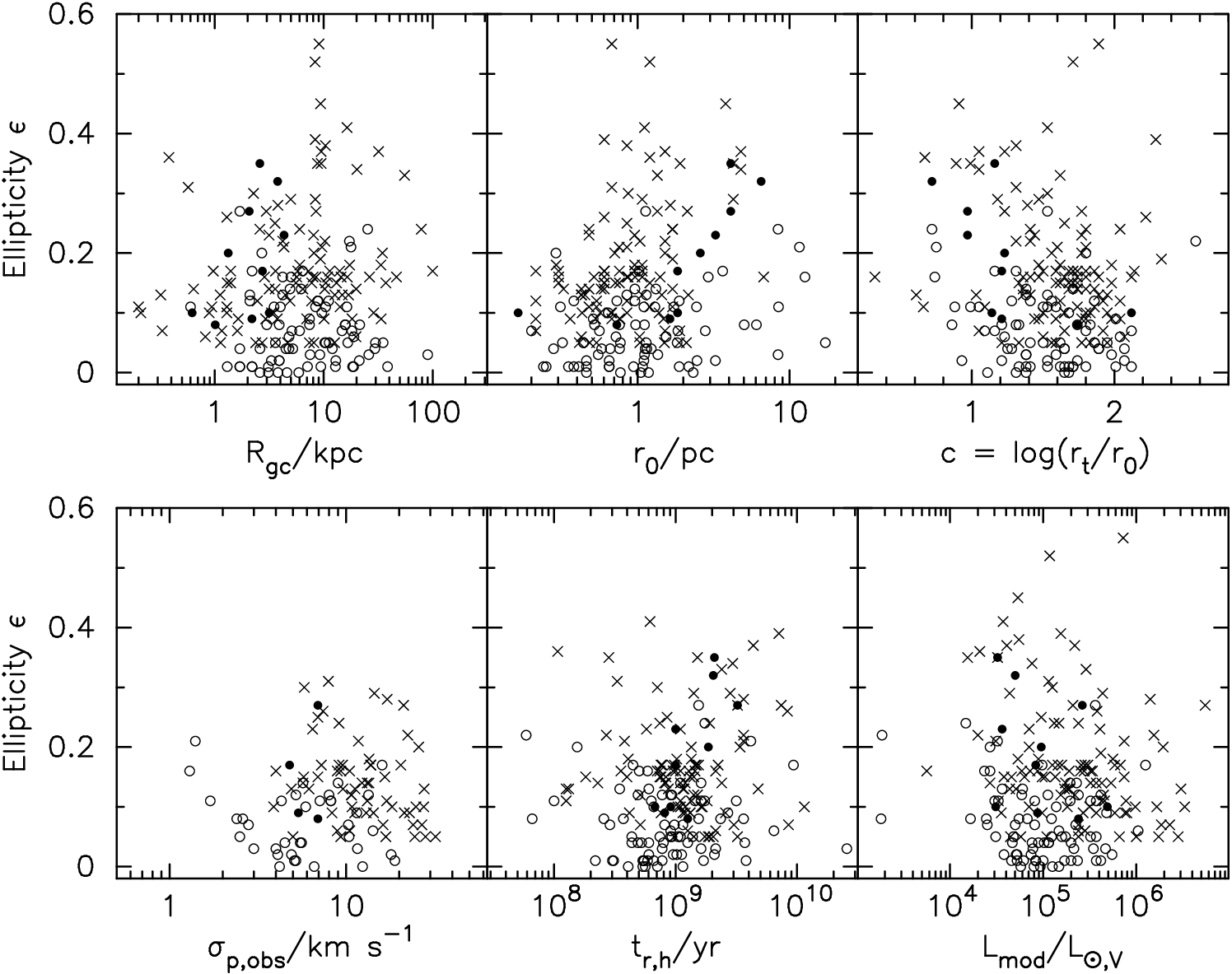}}}
\caption{Ellipticity as a function of galactocentric distance, luminosity, observed velocity dispersion, and some structural parameters. Symbols are as in Fig. 19.}
\label{fig20}
\end{figure*}

\subsection{Ellipticity}

Figure 20 shows the distribution of ellipticity with galactocentric distance, luminosity, observed velocity dispersion, and some structural parameters for clusters in M33, M31 and the Milky Way, which may show clues to the primary factor for the elongation of clusters: galaxy tides, rotation and velocity anisotropy, cluster mergers, and ``remnant elongation'' \citep{larsen01,barmby07}. We used the ellipticities of the M33 sample clusters obtained from the F555W band, since the ellipticities of M31 clusters from \citet{barmby07} were obtained from the F555W or F606W band. In addition, we used the ellipticities of M31 clusters obtained from the F555W or F606W band in \citet{wang13}. In fact, \citet{wang13} have presented this distribution for star clusters in M31 and the Milky Way, and discussed in detail. We include the sample clusters of M33 in this distribution. It is evident that all the clusters in the three galaxies show similar trends. In addition, Figure 20 shows that a seeming correlation exists between ellipticity and $R_{\rm gc}$, $r_0$, and $c$ among M33 halo clusters, and that this correlation does not exist among the clusters in M31 and the Milky Way. We argued that this seeming correlation among M33 halo clusters is not true and it results from small number statistic of M33 clusters. In Figure 20, there are only 10 M33 sample clusters. In fact, \citet{roman12} presented the distribution of ellipticity with $R_{\rm gc}$ for 161 star clusters in M33, and did not find a correlation between ellipticity and $R_{\rm gc}$  (See Figure 2 of San Roman et al. 2012). When we include M33 clusters of \citet{roman12} in Figure 20, any correlations between ellipticity and $r_0$ and $c$ among M33 clusters disappear.

\begin{figure*}[!htb]
\figurenum{21}
\center
\resizebox{\width}{!}{\rotatebox{0}
{\includegraphics[width=0.9\textwidth]{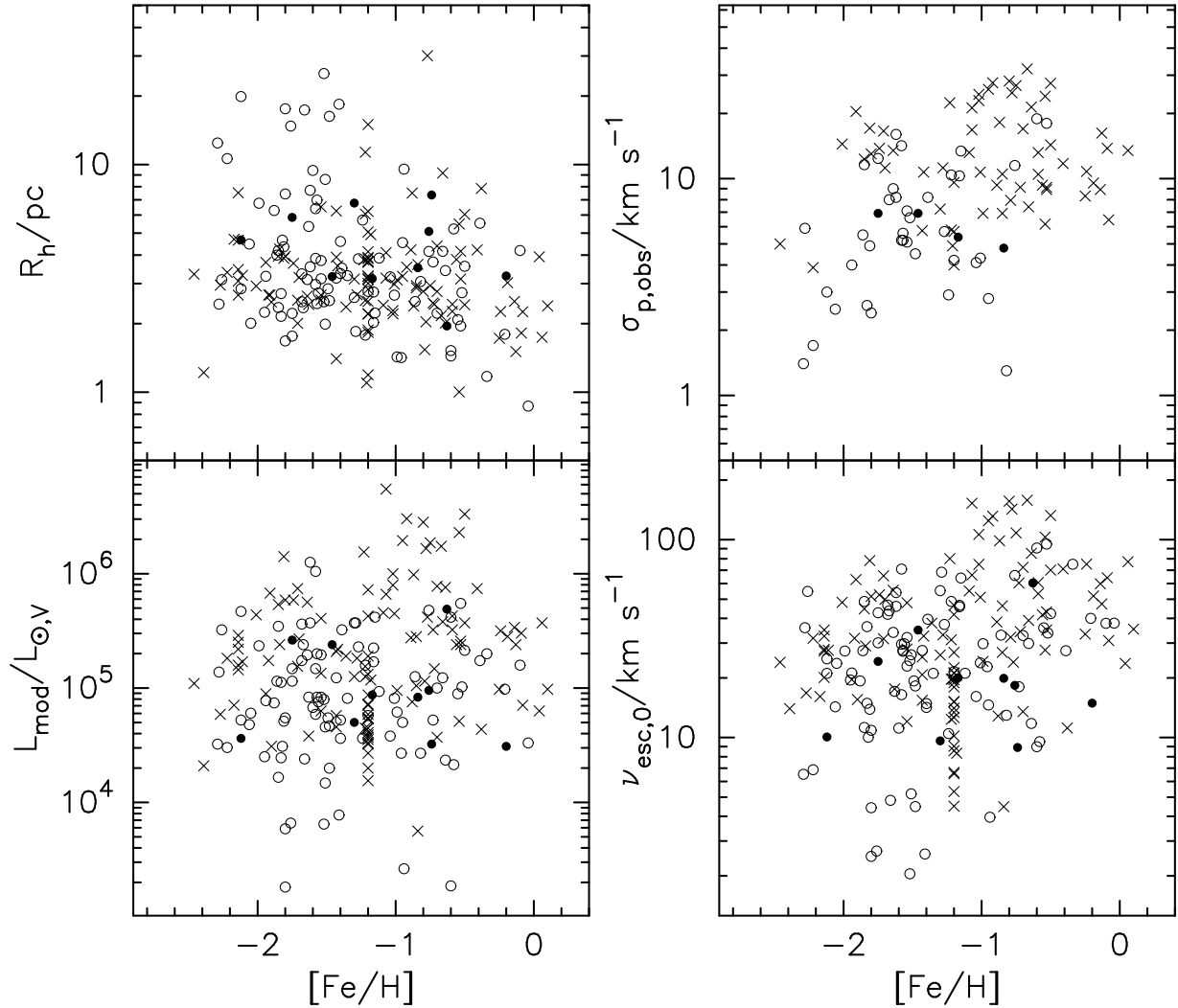}}}
\caption{Structural and dynamical parameters as a function of [Fe/H]. Symbols are as in Fig. 19.}
\label{fig21}
\end{figure*}

\subsection{Metallicity}

Figure 21 plots structural and dynamical parameters as a function of [Fe/H] for clusters in M33, M31 and the Milky Way. The trends of $R_h$, $\sigma_{p,\rm obs}$ and $\nu_{\rm esc,0}$ with [Fe/H] appear. The metal-rich clusters have smaller average values of $R_h$ than those of metal-poor clusters. This result is in agreement with those of \citet{larsen01} and \citet{bhh02}. The metal-rich clusters have larger average values of $\sigma_{p,\rm obs}$ and $\nu_{\rm esc,0}$ than those of metal-poor clusters. However, all the three trends have large scatters. It is true that all the GCs in the three galaxies present similar trends except for the panel of $\sigma_{p,\rm obs}$. In this panel, there are only four sample clusters of M33 having observed $\sigma_{p,\rm obs}$, which have not any correlation or a weak negative correlation with [Fe/H]. However, for the clusters of M31, M33 and the Milky Way together, the observed velocity dispersion $\sigma_{p,\rm obs}$ increases with metallicity. We argued that this seeming contradiction is not true and it results from small number statistic of M33 clusters. In addition, we did not include the data for M31 young massive clusters (YMCs) in Figure 21, since their metallicities are not accurately derived. The solar values of metallicities are adopted for M31 YMCs in \citet{wang13}, who have presented this distribution for clusters in M31 and the Milky Way, and discussed in detail.

\subsection{The Fundamental Plane}

\begin{figure*}[!htb]
\figurenum{22}
\center
\resizebox{\width}{!}{\rotatebox{0}
{\includegraphics[width=0.9\textwidth]{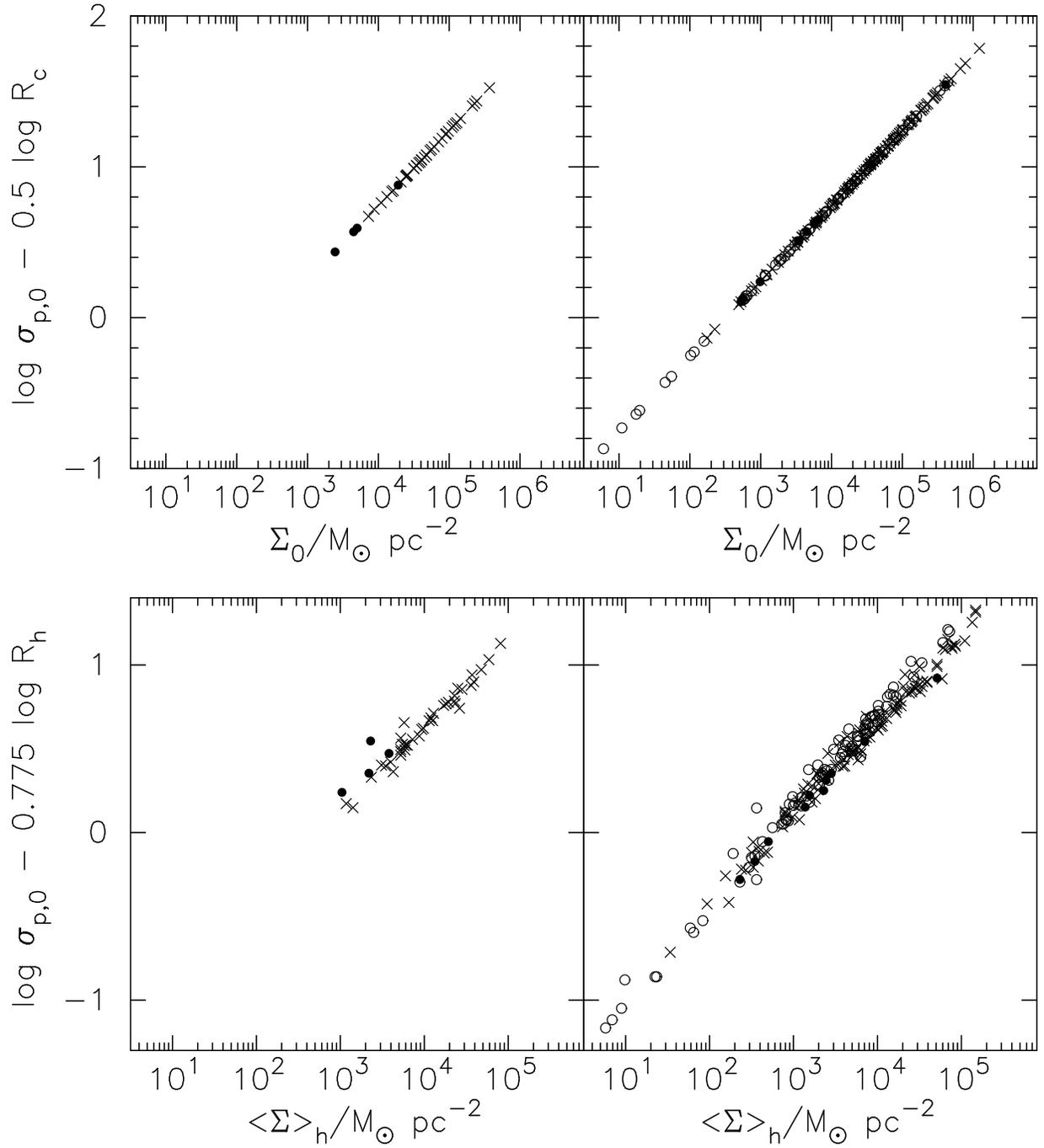}}}
\caption{Evidence of an FP of the cluster parameters, which is defined in terms of central velocity dispersion $\sigma_{p,0}$, projected core radius $R_c$ or projected half-light radius $R_h$, and the central surface mass density $\Sigma_0$ or surface mass density averaged over the half-light radius $\langle\Sigma\rangle_h$. Symbols are as in Fig. 19.}
\label{fig22}
\end{figure*}

\citet{bhh02} demonstrated that both M31 and Milky Way GCs do not occupy the full space of four-dimensional parameters (concentration $c$, scale radius $r_0$, central surface brightness $\mu_{V,0}$, and central $M/L$ or velocity dispersion $\sigma_0$), which describe GC structure. In fact, \citet{dm94} performed a detailed analysis of a data set of cluster parameters on 143 GCs of the Milky Way, and found some monovariate correlations between various properties. Then, \citet{djor95} used the principal component analysis to infer the existence of a pair of bivariate correlations in Milky Way GC parameters which implying the existence of a ``fundamental plane (FP)'', analogous to that of elliptical galaxies. This parameter space includes a radius (core, or half-light), a surface brightness (central, or average within a half-light radius), and the central projected velocity dispersion. However, \citet{mclau00} considered that the statistical analysis cannot offer any physical insight into this FP, and he defined an FP using the binding energy and luminosity, which is formally different from that of \citet{djor95}. \citet{mclau00} derived the relations between $\sigma_{0}$, $r_0$ and $\mu_{V,0}$, and between $\sigma_{0}$, $R_h$ and $\langle\mu_{V,0}\rangle_h$: $\log \sigma_{0} - 0.5 \log r_0 \sim \mu_{V,0}$, and $\log \sigma_{0} - 0.775 \log R_h \sim \langle\mu_{V,0}\rangle_h$ (see eq. [12a] and [13b] of McLaughlin 2000 in detail) based on the basic definitions. Then, \citet{barmby09} presented the correlations between $\sigma_{0}$, $R_c$ and $\mu_{V,0}$ ($\log \sigma_{0} - 0.5 \log R_c$ versus $\mu_{V,0}$), and between $\sigma_{0}$, $R_h$ and $\langle\mu_{V,0}\rangle_h$ ($\log \sigma_{0} - 0.775 \log R_h$ versus $\langle\mu_{V,0}\rangle_h$) and found large offsets between the young M31 clusters and old clusters. However, \citet{barmby09} showed that the correlations between $\sigma_{0}$, $R_c$ and $\Sigma_0$ ($\log \sigma_{0} - 0.5 \log R_c$ versus $\Sigma_0$), and between $\sigma_{0}$, $R_h$ and $\langle\Sigma\rangle_h$ ($\log \sigma_{0} - 0.755 \log R_h$ versus $\langle\Sigma\rangle_h$) are tight, which are called the mass-density-based FP relations. Here we show the two forms of the FP for the clusters in M33, M31 and the Milky Way.

Figure 22 plots the mass-density-based FP relations for the clusters in M33, M31 and the Milky Way. In the left two panels, $\sigma_{p,0}$, $\Sigma_0$ and $\langle\Sigma\rangle_h$ are computed using the observed $M/L$ values. For both M33 and M31 clusters, the observed $M/L$ values are from \citet{larsen02} and \citet{str11}, respectively. In the right two panels, the values of $\sigma_{p,0}$, $\Sigma_0$ and $\langle\Sigma\rangle_h$ are computed using the $M/L$ values from population-synthesis models. It is obvious that the velocity dispersion, characteristic radii, and mass density for M33, M31 and Milky Way clusters show tight correlations, both on the core and half-light scales. The exist of FP relations for clusters strongly reflects some universal physical conditions and processes of cluster formation.

Figure 23 shows the correlation of binding energy with the total model mass for the clusters in M33, M31 and the Milky Way. In the left panel, the values of $M_{\rm mod}$ and $E_b$ are computed using the observed $M/L$ values. For both M33 and M31 clusters, the observed $M/L$ values are from \citet{larsen02} and \citet{str11}, respectively. In the right panel, the values of $M_{\rm mod}$ and $E_b$ are computed using the $M/L$ values from population-synthesis models. It is evident that all the clusters in M33, M31 and the Milky Way locate in a remarkably tight region. \citet{barmby07} concluded that the scatter around this relation is so small that the structures of star clusters may be far simpler than those scenarios derived from theoretical arguments.

\begin{figure*}[!htb]
\figurenum{23}
\center
\resizebox{\width}{!}{\rotatebox{0}
{\includegraphics[width=0.9\textwidth]{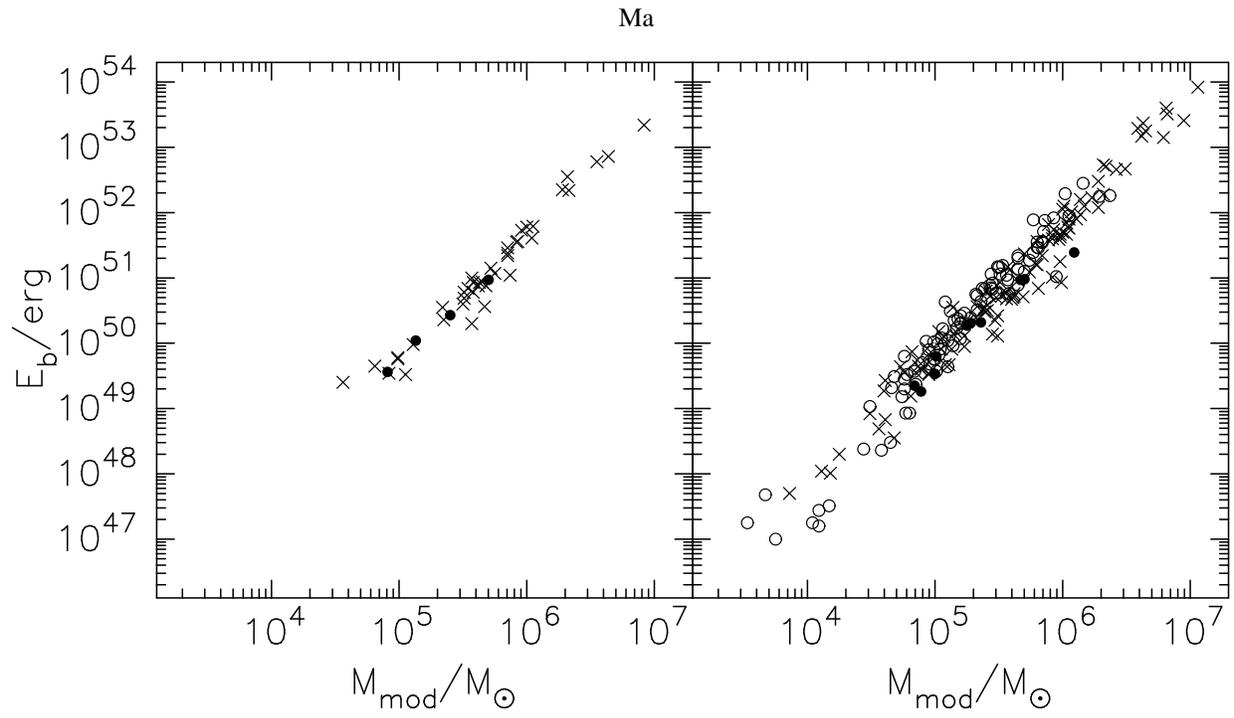}}}
\caption{Evidence of an FP of the cluster parameters, which is defined in terms of binding energy $E_b$ and the total model mass $M_{\rm mod}$. Symbols are as in Fig. 19.}
\label{fig23}
\end{figure*}

\section{SUMMARY}

In this paper, we present the properties of 10 halo GCs in M33 using the {\it HST}/WFPC2 images in the F555W and F814W bands. The sample of M33 halo GCs in this paper is from \citet{sara98}, who originally selected 10 star clusters in M33  based on their halo-like kinematics and red colors ($B-V >0.6$). \citet{sara98} considered that these clusters should be as close an analogy as possible to the halo clusters in the Milky Way. We obtained ellipticities, position angles and surface brightness profiles for them. Structural and dynamical parameters were derived by fitting the profiles to three different models, including King, Wilson, and S\'{e}rsic models. We found that, in the majority of cases, King model fit the M33 clusters as well as Wilson model, and better than S\'{e}rsic model.

We discussed the properties of the sample GCs here combined with GCs in the Milky Way \citep{mm05} and M31 \citep{barmby07,wang13}. In general, the properties of M33, M31 and the Galactic clusters fall in the same regions of parameter spaces. We investigated two forms of the FP, including the correlation of velocity dispersion, radius, and mass density, and the correlation of binding energy with the total model mass. The tight correlations of cluster properties indicate a tight FP for clusters, regardless of their host environments. In addition, the tightness of the relations for the internal properties indicates some physical conditions and processes of cluster formation in different galaxies.

\acknowledgments We would like to thank the anonymous referee for providing rapid and thoughtful report that helped improve the original manuscript greatly. We would like to thank Dr. McLaughlin for his help in deriving the parameters of the three structure models. This work was supported by the National Basic Research Program of China (973 Program), No. 2014CB845702, and by the Chinese National Natural Science Foundation Grands No. 11373035 and 11433005.

\clearpage
\begin{deluxetable*}{cccccccccccc}
\tablecolumns{12}\tablewidth{0pc}
\tablecaption{Integrated Measurements of 10 Halo Globular Clusters in M33\label{table1}}
\tabletypesize{\scriptsize}
\tablehead{\colhead{ID} & \colhead{$\epsilon^{\rm a}$} & \colhead{$\theta^{\rm a}$} & \colhead{$\epsilon^{\rm b}$} & \colhead{$\theta^{\rm b}$} & \colhead{$B^{\rm c}$}  & \colhead{$V^{\rm c}$}  & \colhead{$I^{\rm c}$}  &\colhead{${R_{\rm gc}}^{\rm d}$} &\colhead{$E(B-V)^{\rm e}$}  &\colhead{${\rm [Fe/H]}^{\rm f}$}   &\colhead{${\rm Age}^{\rm g}$}\\
\colhead{}     &  \colhead{}     &  \colhead{(deg E of N)}    & \colhead{}   & \colhead{(deg E of N)}  &  \colhead{(Vegamag)}   & \colhead{(Vegamag)}  &  \colhead{(Vegamag)}  &  \colhead{(kpc)}  &  \colhead{}  & \colhead{} &  \colhead{(Gyr)}\\
\colhead{(1)} &  \colhead{(2)}  &  \colhead{(3)} &  \colhead{(4)} & \colhead{(5)} &  \colhead{(6)}  & \colhead{(7)} &  \colhead{(8)}  &  \colhead{(9)} & \colhead{(10)} &  \colhead{(11)} & \colhead{(12)}}
\startdata
 U49  & $0.27 \pm 0.03 $ & $43   \pm 2    $ & $0.23  \pm 0.02 $ & $40   \pm3    $ & 17.08 & 16.30 & 15.31 & 2.07  & 0.05  & $-1.75 \pm 0.04 $ & $3.98  \pm 0.32 $ \\
 R12  & $0.08 \pm 0.01 $ & $-20  \pm 9    $ & $0.17  \pm 0.02 $ & $10   \pm12   $ & 17.42 & 16.38 & 15.19 & 1.01  & 0.04  & $-1.46 \pm 0.04 $ & $10.00 \pm 0.41 $ \\
 R14  & $0.10 \pm 0.02 $ & $-30  \pm 16   $ & $0.11  \pm 0.01 $ & $-13  \pm2    $ & 17.46 & 16.47 & 15.15 & 0.62  & 0.38  & $-0.63 \pm 0.04 $ & $1.29  \pm 0.23 $ \\
 M9   & $0.09 \pm 0.01 $ & $-15  \pm 10   $ & $0.10  \pm 0.01 $ & $-8   \pm9    $ & 17.88 & 17.13 & 16.11 & 2.19  & 0.03  & $-1.17 \pm 0.11 $ & $4.27  \pm 0.37 $ \\
 U77  & $0.20 \pm 0.02 $ & $0    \pm 17   $ & $0.16  \pm 0.02 $ & $-63  \pm4    $ & 18.04 & 17.29 & 16.38 & 1.33  & 0.06  & $-0.76 \pm 0.05 $ & $1.58  \pm 0.25 $ \\
 H38  & $0.17 \pm 0.03 $ & $26   \pm 22   $ & $0.22  \pm 0.04 $ & $25   \pm17   $ & 18.13 & 17.29 & 16.28 & 2.74  & 0.03  & $-0.84 \pm 0.04 $ & $5.01  \pm 0.16 $ \\
 C20  & $0.32 \pm 0.01 $ & $-32  \pm 12   $ & $0.21  \pm 0.05 $ & $-56  \pm3    $ & 18.43 & 17.58 & 16.59 & 3.77  & 0.02  & $-1.30 \pm 0.07 $ & $8.91  \pm 0.25 $ \\
 C38  & $0.23 \pm 0.04 $ & $-36  \pm 27   $ & $0.25  \pm 0.03 $ & $-25  \pm12   $ & 18.74 & 18.04 & 17.21 & 4.32  & 0.03  & $-2.12 \pm 0.11 $ & $1.91  \pm 0.05 $ \\
\enddata
\tablenotetext{a}{$\epsilon$ and $\theta$ are ellipticity and position angle of F555W filter for each cluster, respectively.}
\tablenotetext{b}{$\epsilon$ and $\theta$ are ellipticity and position angle of F814W filter for each cluster, respectively.}
\tablenotetext{c}{$B$, $V$, and $I$ are the magnitudes in the $B$, $V$, and $I$ bands for each cluster, respectively.}
\tablenotetext{d}{$R_{\rm gc}$ is the galactocentric distance of each cluster.}
\tablenotetext{e}{$E(B-V)$ is the reddening value of each cluster.}
\tablenotetext{f}{${\rm [Fe/H]}$ is the metallicity of each cluster.}
\tablenotetext{g}{${\rm Age}$ is the age of each cluster.}
\end{deluxetable*}

\begin{deluxetable*}{ccccccc}
\tablecolumns{12}\tablewidth{0pc}
\tablecaption{Intensity Profiles for 10 Halo Globular Clusters in M33\label{table2}}
\tablehead{\colhead{ID} &  \colhead{Band} & \colhead{$R^{\rm a}$} & \colhead{$I$} & \colhead{Uncertainty}  & \colhead{Flag} \\
\colhead{}    & \colhead{}   & \colhead{(arcsec)}  &  \colhead{($L_{\odot}$~pc$^{-2}$)}   & \colhead{($L_{\odot}$~pc$^{-2}$)}  &  \colhead{}  \\
\colhead{(1)} &  \colhead{(2)}  &  \colhead{(3)} &  \colhead{(4)} & \colhead{(5)} &  \colhead{(6)}}
\startdata
 U49      & F555W    & 0.0234     & 2047.989     & 2.150        & OK    \\
          & F555W    & 0.0258     & 2047.598     & 2.260        & DEP   \\
          & F555W    & 0.0284     & 2047.044     & 2.332        & DEP   \\
          & F555W    & 0.0312     & 2046.346     & 2.379        & DEP   \\
          & F555W    & 0.0343     & 2045.563     & 2.419        & DEP   \\
          & F555W    & 0.0378     & 2044.743     & 2.416        & DEP   \\
          & F555W    & 0.0415     & 2043.878     & 2.411        & DEP   \\
          & F555W    & 0.0457     & 2042.969     & 2.398        & DEP   \\
          & F555W    & 0.0503     & 2041.900     & 2.332        & OK    \\
          & F555W    & 0.0553     & 2040.737     & 2.283        & DEP   \\
          & F555W    & 0.0608     & 2039.297     & 2.229        & DEP   \\
          & F555W    & 0.0669     & 2037.591     & 2.165        & DEP   \\
          & F555W    & 0.0736     & 2035.377     & 2.115        & OK    \\
\enddata
\tablenotetext{a}{$R$ is the clustercentric radius.}
\tablenotetext{1}{This table is available in its entirely in machine-readable.}
\end{deluxetable*}

\clearpage
\begin{deluxetable*}{cccccrcccrr}
\tablecolumns{12}\tablewidth{0pc}
\tablecaption{Basic Parameters of 10 Halo Globular Clusters in M33\label{table3}}
\tabletypesize{\scriptsize}
\tablehead{\colhead{ID} & \colhead{Band} & \colhead{$N_{\rm pts}$$^{\rm a}$} & \colhead{Model} & \colhead{$\chi_{\rm min}^2$$^{\rm b}$}  & \colhead{$I_{\rm bkg}$$^{\rm c}$}  & \colhead{$W_0$$^{\rm d}$}  &\colhead{$c/n$$^{\rm e}$} &\colhead{$\mu_0$$^{\rm f}$}  &\colhead{$\log r_0$$^{\rm g}$}   &\colhead{$\log r_0$$^{\rm h}$}\\
\colhead{}     & \colhead{}   & \colhead{}  &  \colhead{}   & \colhead{}  &
\colhead{($L_{\odot}$~pc$^{-2}$)}  &  \colhead{}  &  \colhead{}  & \colhead{${\rm (mag~arcsec^{-2})}$} & \colhead{(arcsec)} & \colhead{(pc)}\\
\colhead{(1)} &  \colhead{(2)}  &  \colhead{(3)} &  \colhead{(4)} & \colhead{(5)} &  \colhead{(6)}  & \colhead{(7)} &  \colhead{(8)}  &  \colhead{(9)} & \colhead{(10)} &  \colhead{(11)}}
\startdata
 U49      &F555W      & 47       & K66      & 7.58     & $0.10     \pm 0.75    $ & $4.70    ^{+0.02    }_{-0.01   }         $ & $0.97    ^{+0.01    }_{-0.01   }         $ & $17.95   ^{+0.01    }_{-0.01   }$ & $0.000   ^{+0.009   }_{-0.007  }$ & $0.613   ^{+0.009   }_{-0.007  }$\\
          &           & 47       & W        & 8.82     & $1.90     \pm 1.27    $ & $4.70    ^{+0.01    }_{-0.03   }         $ & $1.35    ^{+0.01    }_{-0.01   }         $ & $17.95   ^{+0.01    }_{-0.01   }$ & $0.000   ^{+0.012   }_{-0.022  }$ & $0.613   ^{+0.012   }_{-0.022  }$\\
          &           & 47       & S        & 58.47    & $2.40     \pm 2.62    $ & $\ldots                                  $ & $0.75    ^{+0.02    }_{-0.01   }         $ & $17.92   ^{+0.04    }_{-0.01   }$ & $-0.050  ^{+0.034   }_{-0.041  }$ & $0.563   ^{+0.034   }_{-0.041  }$\\
          &F814W      & 48       & K66      & 16.01    & $-2.10    \pm 1.12    $ & $5.80    ^{+0.01    }_{-0.02   }         $ & $1.21    ^{+0.01    }_{-0.01   }         $ & $16.94   ^{+0.01    }_{-0.01   }$ & $-0.100  ^{+0.034   }_{-0.030  }$ & $0.513   ^{+0.034   }_{-0.030  }$\\
          &           & 48       & W        & 27.99    & $2.00     \pm 3.53    $ & $4.60    ^{+0.03    }_{-0.05   }         $ & $1.32    ^{+0.01    }_{-0.01   }         $ & $16.95   ^{+0.01    }_{-0.01   }$ & $0.000   ^{+0.040   }_{-0.054  }$ & $0.613   ^{+0.040   }_{-0.054  }$\\
          &           & 48       & S        & 26.77    & $-0.50    \pm 0.85    $ & $\ldots                                  $ & $1.00    ^{+0.01    }_{-0.01   }         $ & $18.63   ^{+0.20    }_{-0.18   }$ & $-0.200  ^{+0.030   }_{-0.036  }$ & $0.413   ^{+0.030   }_{-0.036  }$\\
 R12      &F555W      & 49       & K66      & 6.32     & $4.10     \pm 1.06    $ & $7.70    ^{+0.05    }_{-0.16   }         $ & $1.74    ^{+0.02    }_{-0.05   }         $ & $15.74   ^{+0.02    }_{-0.05   }$ & $-0.750  ^{+0.018   }_{-0.015  }$ & $-0.137  ^{+0.018   }_{-0.015  }$\\
\enddata
\tablenotetext{a}{The number of points in the intensity profile that were used for constraining the model fits.}
\tablenotetext{b}{The minimum $\chi^2$ obtained in the fits.}
\tablenotetext{c}{The best-fitted background intensity.}
\tablenotetext{d}{The dimensionless central potential of the best-fitting model, defined as $W_0 \equiv -\phi(0)/\sigma_0^2$.}
\tablenotetext{e}{The concentration $c \equiv \log(r_t/r_0)$.}
\tablenotetext{f}{The best-fit central surface brightness in the native band of the data.}
\tablenotetext{g}{The best model-fit scale radius $r_0$ in arcseconds.}
\tablenotetext{h}{The best model-fit scale radius $r_0$ in parsecs.}
\end{deluxetable*}

\setcounter{table}{3}
\begin{sidewaystable}
\begin{center}
\small
\caption{Derived Structural and Photometric Parameters of 10 Halo Globular Clusters in M33} \vspace{0.0mm} \label{table4}
\tabcolsep=2.5pt
\begin{tabular}{ccccrcccccccc}
\hline\hline
ID & Band & Model & $\log r_{\rm tid}$$^{\rm a}$  & $\log R_c$$^{\rm b}$  & $\log R_h$$^{\rm c}$  &$\log R_h/R_c$$^{\rm d}$ &$\log I_{\rm 0}$$^{\rm e}$  & $\log j_{\rm 0}$$^{\rm f}$  & $\log L_V$$^{\rm g}$ & $V_{\rm tot}$$^{\rm h}$ & $\log I_h$$^{\rm i}$ & $\langle\mu_V\rangle_h$$^{\rm j}$\\
     &    &   &  (pc)   & (pc)  &
(pc)  &    &  ($L_{\odot,V}$~pc$^{-2}$) & ($L_{\odot,V}$~pc$^{-3}$) &  ($L_{\odot,V}$) &   (mag) &   ($L_{\odot,V}$~pc$^{-2}$) &   ${\rm (mag~arcsec^{-2})}$\\
(1) &  (2)  &  (3) &  (4) & (5) &  (6)  & (7) &  (8)  &  (9) & (10) &  (11) & (12) &  (13)\\
\hline
 U49      & F555W       & K66      & $1.58    ^{+0.01    }_{-0.01   }         $ & $0.552   ^{+0.009   }_{-0.007  }$ & $0.768   ^{+0.010   }_{-0.010  }$ & $0.216   ^{+0.017   }_{-0.019  }$ & $3.38    ^{+0.02    }_{-0.02   }$ & $2.52    ^{+0.03    }_{-0.03   }$ & $5.42    ^{+0.02    }_{-0.02   }$ & $15.86   ^{+0.05    }_{-0.05   }$ & $3.09    ^{+0.01    }_{-0.01   }$ & $18.63   ^{+0.01    }_{-0.01   }$\\
          &             & W        & $1.96    ^{+0.01    }_{-0.01   }         $ & $0.528   ^{+0.011   }_{-0.022  }$ & $0.776   ^{+0.010   }_{-0.010  }$ & $0.248   ^{+0.032   }_{-0.022  }$ & $3.38    ^{+0.02    }_{-0.02   }$ & $2.54    ^{+0.04    }_{-0.03   }$ & $5.41    ^{+0.02    }_{-0.02   }$ & $15.91   ^{+0.05    }_{-0.05   }$ & $3.06    ^{+0.01    }_{-0.01   }$ & $18.72   ^{+0.01    }_{-0.01   }$\\
\hline
\end{tabular}
\end{center}
\end{sidewaystable}

\setcounter{table}{3}
\begin{sidewaystable}
\begin{center}
\small
\caption{(Continued)} \vspace{0.0mm} \label{table4}
\tabcolsep=2.5pt
\begin{tabular}{ccccrcccccccc}
\hline\hline
ID & Band & Model & $\log r_{\rm tid}$$^{\rm a}$  & $\log R_c$$^{\rm b}$  & $\log R_h$$^{\rm c}$  &$\log R_h/R_c$$^{\rm d}$ &$\log I_{\rm 0}$$^{\rm e}$  & $\log j_{\rm 0}$$^{\rm f}$  & $\log L_V$$^{\rm g}$ & $V_{\rm tot}$$^{\rm h}$ & $\log I_h$$^{\rm i}$ & $\langle\mu_V\rangle_h$$^{\rm j}$\\
     &    &   &  (pc)   & (pc)  &
(pc)  &    &  ($L_{\odot,V}$~pc$^{-2}$) & ($L_{\odot,V}$~pc$^{-3}$) &  ($L_{\odot,V}$) &   (mag) &   ($L_{\odot,V}$~pc$^{-2}$) &   ${\rm (mag~arcsec^{-2})}$\\
(1) &  (2)  &  (3) &  (4) & (5) &  (6)  & (7) &  (8)  &  (9) & (10) &  (11) & (12) &  (13)\\
\hline
 C38      & F555W       & K66      & $1.48    ^{+0.01    }_{-0.02   }         $ & $0.452   ^{+0.027   }_{-0.049  }$ & $0.669   ^{+0.010   }_{-0.010  }$ & $0.218   ^{+0.059   }_{-0.037  }$ & $2.71    ^{+0.02    }_{-0.02   }$ & $1.95    ^{+0.07    }_{-0.05   }$ & $4.56    ^{+0.02    }_{-0.02   }$ & $18.03   ^{+0.05    }_{-0.05   }$ & $2.42    ^{+0.01    }_{-0.01   }$ & $20.30   ^{+0.01    }_{-0.01   }$\\
          &             & W        & $1.81    ^{+0.02    }_{-0.01   }         $ & $0.421   ^{+0.044   }_{-0.040  }$ & $0.658   ^{+0.020   }_{-0.009  }$ & $0.237   ^{+0.060   }_{-0.053  }$ & $2.71    ^{+0.02    }_{-0.02   }$ & $1.98    ^{+0.06    }_{-0.07   }$ & $4.52    ^{+0.02    }_{-0.02   }$ & $18.12   ^{+0.05    }_{-0.05   }$ & $2.41    ^{+0.01    }_{-0.02   }$ & $20.33   ^{+0.05    }_{-0.01   }$\\
\hline
\end{tabular}
\end{center}
\end{sidewaystable}

\setcounter{table}{4}
\begin{sidewaystable}
\begin{center}
\small
\caption{Derived Dynamical Parameters of 10 Halo Globular Clusters in M33} \vspace{0.0mm} \label{table5}
\tabcolsep=2.5pt
\begin{tabular}{ccccccccccccr}
\hline\hline
ID & Band & $\Upsilon_V^{\rm pop}$$^{\rm a}$ & Model  & $\log M_{\rm tot}$$^{\rm b}$  & $\log E_b$$^{\rm c}$  & $\log \Sigma_{\rm 0}$$^{\rm d}$ & $\log \rho_{\rm 0}$$^{\rm e}$  &$\log \Sigma_h$$^{\rm f}$   &$\log \sigma_{p,0}$$^{\rm g}$ & $\log \nu_{\rm esc,0}$$^{\rm h}$ & $\log t_{r,h}$$^{\rm i}$ & $\log f_{\rm 0}$$^{\rm j}$\\
     &    & ($M_{\odot}~L_{\odot,V}^{-1}$)  &     & ($M_{\odot}$)  &
(erg)  &  ($M_{\odot}$~pc$^{-2}$)  &  ($M_{\odot}$~pc$^{-3}$) & ($M_{\odot}$~pc$^{-2}$) &  (km s$^{-1}$) &   (km s$^{-1}$) &   (yr) &   ($M_{\odot}$~(pc~km~s$^{-1})^{-3})$\\
(1) &  (2)  &  (3) &  (4) & (5) &  (6)  & (7) &  (8)  &  (9) & (10) &  (11) & (12) &  (13)\\
\hline
 U49      &F555W       & $1.879   ^{+0.253   }_{-0.213  }         $ & K66      & $5.70    ^{+0.06    }_{-0.06   }$ & $50.98   ^{+0.22    }_{-0.21   }$ & $3.65    ^{+0.06    }_{-0.06   }$ & $2.79    ^{+0.06    }_{-0.06   }$ & $3.36    ^{+0.05    }_{-0.05   }$ & $0.846   ^{+0.028   }_{-0.027  }$ & $1.385   ^{+0.028   }_{-0.027  }$ & $9.51    ^{+0.06    }_{-0.06   }$ & $-0.999  ^{+0.028   }_{-0.029  }$\\
          &            & $                                        $ & W        & $5.68    ^{+0.06    }_{-0.06   }$ & $50.92   ^{+0.22    }_{-0.21   }$ & $3.65    ^{+0.06    }_{-0.06   }$ & $2.81    ^{+0.07    }_{-0.06   }$ & $3.33    ^{+0.05    }_{-0.05   }$ & $0.833   ^{+0.027   }_{-0.026  }$ & $1.395   ^{+0.027   }_{-0.027  }$ & $9.51    ^{+0.06    }_{-0.06   }$ & $-0.943  ^{+0.052   }_{-0.032  }$\\
\end{tabular}
\end{center}
\end{sidewaystable}

\setcounter{table}{4}
\begin{sidewaystable}
\begin{center}
\small
\caption{(Continued)} \vspace{0.0mm} \label{table5}
\tabcolsep=2.5pt
\begin{tabular}{ccccccccccccr}
\hline\hline
ID & Band & $\Upsilon_V^{\rm pop}$$^{\rm a}$ & Model  & $\log M_{\rm tot}$$^{\rm b}$  & $\log E_b$$^{\rm c}$  & $\log \Sigma_{\rm 0}$$^{\rm d}$ & $\log \rho_{\rm 0}$$^{\rm e}$  &$\log \Sigma_h$$^{\rm f}$   &$\log \sigma_{p,0}$$^{\rm g}$ & $\log \nu_{\rm esc,0}$$^{\rm h}$ & $\log t_{r,h}$$^{\rm i}$ & $\log f_{\rm 0}$$^{\rm j}$\\
     &    & ($M_{\odot}~L_{\odot,V}^{-1}$)  &     & ($M_{\odot}$)  &
(erg)  &  ($M_{\odot}$~pc$^{-2}$)  &  ($M_{\odot}$~pc$^{-3}$) & ($M_{\odot}$~pc$^{-2}$) &  (km s$^{-1}$) &   (km s$^{-1}$) &   (yr) &   ($M_{\odot}$~(pc~km~s$^{-1})^{-3})$\\
(1) &  (2)  &  (3) &  (4) & (5) &  (6)  & (7) &  (8)  &  (9) & (10) &  (11) & (12) &  (13)\\
\hline
 C38      &F555W       & $1.914   ^{+0.245   }_{-0.235  }         $ & K66      & $4.84    ^{+0.06    }_{-0.06   }$ & $49.35   ^{+0.23    }_{-0.23   }$ & $2.99    ^{+0.06    }_{-0.06   }$ & $2.23    ^{+0.09    }_{-0.07   }$ & $2.70    ^{+0.05    }_{-0.06   }$ & $0.464   ^{+0.030   }_{-0.029  }$ & $1.003   ^{+0.029   }_{-0.028  }$ & $9.00    ^{+0.06    }_{-0.06   }$ & $-0.418  ^{+0.115   }_{-0.066  }$\\
          &            & $                                        $ & W        & $4.81    ^{+0.06    }_{-0.06   }$ & $49.28   ^{+0.22    }_{-0.24   }$ & $3.00    ^{+0.06    }_{-0.06   }$ & $2.26    ^{+0.08    }_{-0.09   }$ & $2.69    ^{+0.06    }_{-0.06   }$ & $0.452   ^{+0.028   }_{-0.031  }$ & $1.012   ^{+0.027   }_{-0.030  }$ & $8.97    ^{+0.07    }_{-0.06   }$ & $-0.353  ^{+0.094   }_{-0.104  }$\\
\hline
\end{tabular}
\end{center}
\end{sidewaystable}

\begin{table}
\centering
\caption{Galactocentric Radii and $\kappa-$Space Parameters of 10 Halo Globular Clusters in M33} \vspace{0.0mm} \label{table6}
\begin{tabular}{ccccrcr}
\hline\hline
\tabletypesize{\scriptsize}
ID & Band & Model & $R_{\rm gc}$ & $\kappa_{m,1}$ & $\kappa_{m,2}$  & $\kappa_{m,3}$ \\
 &    &   &  (kpc) &  &    & \\
(1) &  (2)  &  (3) &  (4) & (5) &  (6)  & (7)\\
\hline
 U49      &F555W       & K66      & 2.1      & $-0.382  ^{+0.042   }_{-0.040  }$ & $4.348   ^{+0.067   }_{-0.064  }$ & $0.323   ^{+0.001   }_{-0.001  }$\\
          &            & W        &          & $-0.395  ^{+0.039   }_{-0.039  }$ & $4.307   ^{+0.067   }_{-0.064  }$ & $0.322   ^{+0.007   }_{-0.001  }$\\
          &            & S        &          & $-0.414  ^{+0.043   }_{-0.043  }$ & $4.388   ^{+0.088   }_{-0.086  }$ & $0.297   ^{+0.005   }_{-0.001  }$\\
          &F814W       & K66      & 2.1      & $-0.420  ^{+0.039   }_{-0.037  }$ & $4.265   ^{+0.068   }_{-0.066  }$ & $0.292   ^{+0.011   }_{-0.013  }$\\
          &            & W        &          & $-0.404  ^{+0.042   }_{-0.037  }$ & $4.326   ^{+0.070   }_{-0.065  }$ & $0.315   ^{+0.024   }_{-0.017  }$\\
\hline
\end{tabular}
\end{table}
\end{document}